\documentclass[11pt]{article}
\usepackage{jheppub}
\usepackage{bm,bbm}
\usepackage{amsmath}
\usepackage{graphics}
\usepackage{braket}
\usepackage{tikz}
\usepackage{mathtools}
\usepackage{comment}
\usepackage{autobreak}
\usepackage{subcaption}

\usepackage{booktabs}

\usetikzlibrary{patterns}
\usetikzlibrary{decorations.markings}
\usetikzlibrary{decorations.pathmorphing}
\tikzset{snake it/.style={decorate, decoration=snake}}

\tikzset{middlearrow/.style={
        decoration={markings,
            mark= at position 0.5 with {\arrow{#1}} ,
        },
        postaction={decorate}
    }
}

\edef\restoreparindent{\parindent=\the\parindent\relax}
\usepackage{parskip}
\restoreparindent
\usepackage{ascmac}
\usepackage{mathrsfs}

\def\Tr{{\rm Tr}}

\def\d{{\rm d}}

\def\i{{\rm i}}

\def\CA{{\cal A}}

\def\CK{{\cal K}}

\def\CR{{\cal R}}
\def\CM{{\cal M}}

\def\BZ{\mathbb{Z}}

\def\b0{\bm{0}_\perp}

\DeclareFontFamily{U}{mathx}{\hyphenchar\font45}
\DeclareFontShape{U}{mathx}{m}{n}{
      <5> <6> <7> <8> <9> <10>
      <10.95> <12> <14.4> <17.28> <20.74> <24.88>
      mathx10
      }{}
\DeclareSymbolFont{mathx}{U}{mathx}{m}{n}
\DeclareFontSubstitution{U}{mathx}{m}{n}
\DeclareMathAccent{\widecheck}{0}{mathx}{"71}
\title{Replica wormholes and capacity of entanglement}

\author[a]{Kohki Kawabata,}
\author[b]{Tatsuma Nishioka,}
\author[a,b]{Yoshitaka Okuyama}
\author[c]{and Kento Watanabe}

\affiliation[a]{Department of Physics, Faculty of Science,
The University of Tokyo, \\
Bunkyo-ku, Tokyo 113-0033, Japan}
\affiliation[b]{Yukawa Institute for Theoretical Physics, Kyoto University,\\
Kitashirakawa Oiwakecho, Sakyo-ku, Kyoto 606-8502, Japan}
\affiliation[c]{Center for Quantum Mathematics and Physics (QMAP), \\
Department of Physics, University of California, Davis, CA 95616 USA}

\abstract{
We consider the capacity of entanglement as a probe of the Hawking radiation in a two-dimensional dilaton gravity coupled with conformal matter of large degrees of freedom.
A formula calculating the capacity is derived using the gravitational path integral, from which we speculate that the capacity has a discontinuity at the Page time in contrast to the continuous behavior of the generalized entropy.
We apply the formula to a replica wormhole solution in an eternal AdS black hole coupled to a flat non-gravitating bath and show that the capacity of entanglement is saturated by the thermal capacity of the black hole in the high temperature limit.
}

\begin{document}

\preprint{YITP-21-45}

\maketitle
\section{Introduction}
The last couple of years has seen considerable progress in the understanding of the black hole information problem \cite{Hawking:1976ra} which poses a tension between the unitary time evolution of black hole and the independence of the Hawking radiation on the initial state of matter \cite{Hawking:1974sw}. 
Recently \cite{Penington:2019npb,Almheiri:2019psf} addressed the problem for a black hole in two-dimensional Anti-de Sitter space (AdS$_2$) attached to an auxiliary system in Minkowski space, which is further justified in the doubly holographic realization \cite{Almheiri:2019hni} 
(for an introductory review see \cite{Almheiri:2020cfm}).
A key to the resolution for the problem is called the island formula \cite{Almheiri:2019hni} 
\begin{align}\label{island_formula_intro}
    S(\text{Rad}) =\underset{\text{I}}{\min}\left[
    \underset{\text{I}}{\mathrm{ext}}\left[
    \frac{\text{Area} (\partial \text{I})}{4G_N}
    + S_\text{mat}(\text{Rad}\cup \text{I})\right]
    \right]\ ,
\end{align}
that accounts for the Page curve for a unitary black hole evaporation process \cite{Page:1993df,Page:1993wv} by incorporating a region $\text{I}$ (island) inside the black hole into a region $\text{Rad}$ (radiation or thermal bath) outside where we measure the (entanglement) entropy of the Hawking radiation.
This is seen as a generalization of the Ryu-Takayanagi formula \cite{Ryu:2006bv,Ryu:2006ef,Hubeny:2007xt,Faulkner:2013ana,Engelhardt:2014gca} for holographic entanglement entropy which derives from the gravitational path integral on the replica spacetime \cite{Lewkowycz:2013nqa,Dong:2016fnf,Dong:2016hjy}, and the island region $\text{I}$ arises from replica wormholes connecting the interior of the black hole in the gravitational path integral picture \cite{Penington:2019kki,Almheiri:2019qdq} (see also \cite{Goto:2020wnk,Colin-Ellerin:2020mva,Colin-Ellerin:2021}). The virtue of the formula is that, once we believe it, we do not have to worry about replica calculation in the same way as the Ryu-Takayanagi formula.

The island formula has been applied to a wide range of black holes with and without evaporation and shown to yield the Page curves consistent with their unitary evolution 
in various backgrounds such as asymptotic flat spaces  \cite{Gautason:2020tmk,Anegawa:2020ezn,Hashimoto:2020cas,Hartman:2020swn,Krishnan:2020oun,Matsuo:2020ypv,Wang:2021woy}, 
de-Sitter spaces \cite{Hartman:2020khs,Balasubramanian:2020xqf,Sybesma:2020fxg,Geng:2021wcq,Aalsma:2021bit}, 
two disjoint universes \cite{Balasubramanian:2020coy,Miyata:2021ncm,Balasubramanian:2021wgd}, 
gravitating thermal baths \cite{Geng:2020fxl,Anderson:2021vof,Balasubramanian:2021wgd},  
shock-wave excited geometries  \cite{Goto:2020wnk,Hollowood:2020cou,Hollowood:2020kvk} and 
 extended theories of gravity \cite{Alishahiha:2020qza}. 
By considering holographic matter, the formula has been justified from the doubly holographic models \cite{Almheiri:2019hni,Almheiri:2019yqk,Rozali:2019day,Chen:2019uhq,Almheiri:2019psy,Balasubramanian:2020hfs,Chen:2020uac,Bak:2020enw,Bousso:2020kmy,Geng:2020qvw,Krishnan:2020fer,Chen:2020jvn,Chen:2020hmv,Ling:2020laa,Hernandez:2020nem,Caceres:2020jcn,Verheijden:2021yrb,Bhattacharya:2021jrn,Ghosh:2021axl,Uhlemann:2021nhu,Geng:2021iyq}
combining the AdS/BCFT models \cite{Takayanagi:2011zk,Fujita:2011fp} and the brane world holography \cite{Randall:1999ee,Randall:1999vf,Karch:2000ct} 
(see also 
\cite{Akal:2020wfl,Miao:2020oey,Miao:2021ual} for wedge holography). 
Moreover islands have been examined through other measures including entanglement negativity \cite{Basak:2020aaa}, reflected entropy \cite{Chandrasekaran:2020qtn,Li:2020ceg}, relative entropy \cite{Chen:2020ojn}, complexity \cite{Zhao:2019nxk,Bhattacharya:2020uun,Hernandez:2020nem,Bhattacharya:2021jrn}, and make manifest themselves and play a central role in baby universes  \cite{Akers:2019nfi,Marolf:2020xie,Giddings:2020yes,Marolf:2020rpm}, global symmetry violation \cite{Harlow:2020bee,Chen:2020ojn,Hsin:2020mfa},
bra-ket wormholes \cite{Chen:2020tes,Numasawa:2020sty}, moving mirror models \cite{Akal:2020twv,Kawabata:2021hac,Reyes:2021npy} (see also 
\cite{Nomura:2019qps,Chen:2019iro,Suzuki:2019xdq,Kusuki:2019hcg,Agon:2020fqs,Dong:2020uxp,Mirbabayi:2020fyk,Akal:2020ujg,Kirklin:2020zic,Liu:2020gnp,Liu:2020jsv,Piroli:2020dlx,Nomura:2020ska,Pollack:2020gfa,Chakravarty:2020wdm,Stanford:2020wkf,Karlsson:2020uga,Engelhardt:2020qpv,Verlinde:2020upt,Murdia:2020iac,Manu:2020tty,Krishnan:2021faa,Bousso:2021sji,Miyaji:2021ktr,Fallows:2021sge,Hollowood:2021nlo,Qi:2021sxb} for related studies).

\renewcommand{\arraystretch}{1.3}
\begin{table}
	\begin{center}
	\begin{tabular}{cc}
	\toprule
	 \textbf{Statistical mechanics}    &\textbf{R{\'e}nyi analogue} \\ \hline
{\footnotesize\begin{tabular}{rl}
     Inverse temperature:& $\displaystyle \beta$    \\
     Hamiltonian:& $H$ \\
     Partition function:& $\displaystyle Z(\beta) = \Tr\left[ e^{-\beta\,H}\right]$ \\ 
     Free energy:   & $\displaystyle F(\beta) = - \beta^{-1}\log Z(\beta)$\\
     Energy: &  $\displaystyle E(\beta) = -\partial_\beta \log Z(\beta)$\\
     Thermal entropy:  & $\displaystyle S(\beta) = \beta^2\, \partial_\beta F(\beta)$\\
     Heat capacity: &  $\displaystyle C(\beta) = - \beta\, \partial_\beta S(\beta)$
\end{tabular}}
& {\footnotesize\begin{tabular}{rl}
     Replica parameter:& $\displaystyle n$\\
     Modular Hamiltonian:& $\displaystyle H_A= -\log\rho_A$\\
     Replica partition function:& $\displaystyle Z(n) = \Tr_A\left[ e^{-n\,H_A}\right]$ \\
     Replica free energy:& $\displaystyle F(n) = - n^{-1}\log Z(n)$\\
     Replica energy:& $\displaystyle E(n) = -\partial_n \log Z(n)$\\
     Refined R{\'e}nyi entropy:& $\displaystyle \tilde{S}^{(n)} = n^2\, \partial_n F(n)$\\
     Capacity of entanglement:&  $\displaystyle C^{(n)} = - n\, \partial_n \tilde{S}^{(n)}$
\end{tabular}}\\
	\bottomrule
	\end{tabular}
	\end{center}
	\caption{An analogy between statistical mechanics and R\'{e}nyi entropic quantities. $\rho_A$ denotes the density matrix associated with a subsystem $A$. Under the identifications $\beta = n$ and $H = H_A$, the entanglement entropy and the capacity of entanglement are equivalent to the thermal entropy and the heat capacity, respectively.}
	\label{Tab:thermal_entanglement}
\end{table}

In the prequel \cite{Kawabata:2021hac} we examined a different quantum information measure than entanglement entropy known as the capacity of entanglement in toy models of black holes with Hawking radiation and proposed that it can be an invaluable probe of the formation of island regions or equivalently the phase transition of dominant replica wormholes.
Originally, the capacity of entanglement was suggested as a useful quantity to characterize topologically ordered states in condensed matter physics \cite{Yao:2010woi,Schliemann_2011}.
It is a quantum informational counterpart of the heat capacity in statistical mechanics as shown in Table \ref{Tab:thermal_entanglement}.
As in statistical mechanics, the capacity of entanglement provides a different kind of information about quantum entanglement than the entanglement entropy.
While there have been considerable amount of works on entanglement entropy, the capacity of entanglement has attracted less attention so far due to its calculational complexity and only a fraction of works has been done in field theories \cite{Nakagawa:2017wis,deBoer:2018mzv,Okuyama:2021ylc} and holography \cite{Nakaguchi:2016zqi}.\footnote{For a certain class of supersymmetric theories there are exact results on the supersymmetric R\'enyi entropy \cite{Nishioka:2013haa} in various dimensions \cite{Nishioka:2014mwa,Huang:2014gca,Hama:2014iea,Alday:2014fsa,Huang:2014pda,Zhou:2015cpa,Zhou:2015kaj,Nian:2015xky,Giveon:2015cgs,Mori:2015bro,Nishioka:2016guu,Yankielowicz:2017xkf,Hosseini:2019and}, for which the corresponding capacity can be defined and calculated in principle.}
This situation prompts us to continue to explore to what extent the capacity of entanglement can discriminate replica wormholes in more realistic setups of an evaporating black hole than those in the previous work \cite{Kawabata:2021hac}.

The main objective of this paper is to derive a formula for the capacity of entanglement applicable for a two-dimensional dilaton gravity coupled to a matter CFT with a large central charge.
We extend the gravitational path integral formulation \cite{Lewkowycz:2013nqa,Dong:2016fnf,Nakaguchi:2016zqi} for the generalized entropy \cite{Almheiri:2019qdq} to higher order in the replica parameter $n$ away from $n=1$, ending up with the formula:
\begin{align}\label{eq:generalcap_intro}
    C =   -\sum_{w_i\in \partial \text{I}} \frac{\partial \Phi^{(n)}(w)}{\partial n}\bigg|_{n=1,w=w_i} + C_{\rm mat}\ ,
\end{align}
where $\Phi^{(n)}$ is the on-shell dilaton field on a spacetime dynamically determined by the equation of motion for fixed $n$.
This formula does not fix the dominant saddle of the gravitational path integral by itself, but should rather be evaluated on a saddle solution follows from the island formula \eqref{island_formula_intro}.
If there is a phase transition between two saddles the entropy is continuous at the Page time when the two saddles switch with each other, whereas the capacity can be discontinuous in general.
Thus it may be phrased in thermodynamic language that the Hawking radiation is a second-order phase transition between the black hole and replica wormhole solutions.

While the entanglement entropy can be determined by the island formula \eqref{island_formula_intro} without knowing the replica geometries other than $n=1$, the capacity of entanglement involves the derivatives of dilatons with respect to $n$ at the conical singularities, which is intrinsically associated with the information beyond the $n=1$ solution.
On the one hand, this is the virtue of the capacity as it can quantify the fluctuation around a dominant saddle in the gravitational path integral, which is inaccessible to the entanglement entropy.
On the other hand, the $n$ dependence prevents us from applying the formula \eqref{eq:generalcap_intro} to the calculation of the capacity for a class of evaporating black holes due to the so-called conformal welding problem \cite{Almheiri:2019hni}, which can be solved numerically \cite{Mirbabayi:2020fyk} but the analytic solutions have been given in very limited cases such as in the high temperature limit.

To circumvent the welding problem we consider an eternal black hole in AdS$_2$ space coupled to a flat bath region in the high temperature limit.
We probe the Hawking radiation by the entanglement entropy for a radiation region extending to the infinity in the bath and show the replica wormhole with an island region has a lower value than the black hole without an island.
Thus the replica wormhole always dominates in the gravitational path integral in the limit and there is no phase transition between dominant saddles.
We then calculate the capacity by applying the formula \eqref{eq:generalcap_intro} to the replica wormhole in the limit, where the $n$-dependence of the dilaton can also be determined analytically.
We find both entanglement entropy and capacity are saturated by the thermal entropy and capacity of the black hole as observed in the previous study \cite{Kawabata:2021hac} for a toy model of radiating black holes.

This paper is organized as follows.
In section \ref{sec:Gravitational path integral in JT gravity} we perform the gravitational path integral derivation of the capacity formula \eqref{eq:generalcap_intro} in a two-dimensional dilaton gravity extending the Jackiw-Teitelboim (JT) gravity coupled to a CFT with a large central charge, and comment on some implications and limitations of the formula.
In section \ref{sec:high temp limit AdS2} we consider an eternal black hole in AdS$_2$ coupled to a flat bath in the high temperature limit where the welding problem becomes trivial and the $n$-dependence of the dilaton can be solved simultaneously.
The replica wormhole solution is shown to always be favored in the island formula \eqref{island_formula_intro} and we evaluate the capacity by applying \eqref{eq:generalcap_intro} to the solution.
We then show both the entanglement entropy and the capacity of entanglement are saturated by the thermal counterparts of the black hole solution in the high temperature limit.
Section \ref{ss:discussion} discusses open problems and future directions.
Two appendices complement some calculations given in the main text.

\section{Gravitational path integral in JT gravity}\label{sec:Gravitational path integral in JT gravity}

Entanglement entropy can be seen as an analogue of thermodynamic entropy at a fixed temperature in the replica trick calculation when the replica parameter $n$ is regarded as the inverse temperature.
Given this analogy, one can introduce the capacity of entanglement as a counterpart of heat capacity in thermodynamics (see e.g.\,\cite{Nakaguchi:2016zqi}).
The capacity of entanglement has potential applications as an indispensable probe of quantum fluctuation of matter and spacetime in various fields \cite{Yao:2010woi,Nakaguchi:2016zqi,Nakagawa:2017wis,deBoer:2018mzv,Verlinde:2019ade}.

In section \ref{ss:renyi}, we expand on the gravitational path integral derivation \cite{Almheiri:2019qdq,Penington:2019kki} of the island formula \eqref{island_formula_intro} in a two-dimensional dilaton gravity coupled to a CFT with a large central charge and derive a similar formula for a refined version of the R\'{e}nyi entropy, which plays a role of thermal entropy at the inverse temperature proportional to $n$.
In section \ref{ss:capacity_ent} we define and derive the capacity of entanglement as a derivative of the refined R\'{e}nyi entropy with respect to $n$ in the gravitational path integral.
Section \ref{sec:disc of CoE} discusses an implication of the capacity formula for the Hawking radiation and we speculate that the capacity typically shows a discontinuity at the Page time in contrast to the entanglement entropy that is continuous.
Section \ref{ss:welding} reviews the conformal welding problem which makes it difficult to calculate the capacity analytically even near $n=1$.

\subsection{R\'{e}nyi entropy}
\label{ss:renyi}
Let us consider a two-dimensional dilaton gravity theory coupled to matter which is assumed to be a two-dimensional CFT with a large central charge. 
In the semiclassical limit,\footnote{When the central charge $c$ satisfies $1 \ll c \ll 1/G_N$, we can decouple the matter theory from the dynamical gravity while keeping the dominant contribution of the gravitational part.}
we have the action:
\begin{align}
    \log Z = -I_{\rm grav} + \log Z_{\rm CFT} \ .
\end{align}
The matter CFT theory lives on the metric $g$ determined by the equation of motion.
For concreteness, we will consider a general dilaton gravity which extends the Jackiw-Teitelboim (JT) gravity \cite{Almheiri:2014cka,Jackiw:1984je,Teitelboim:1983ux} and CGHS model \cite{Callan:1992rs}:
\begin{align}\label{eq:action}
    \begin{aligned}
    I_{\rm grav} &= I_{\rm dil} + I_{\rm EH} + I_{\rm bdy} \ , \\
    I_{\rm dil} &=- \frac{1}{16\pi G_N}\int_{\Sigma_2}
    \Phi\, \left[\CR+U(\Phi)(\nabla \Phi)^2 + V(\Phi)\right]
     \ , \\
    I_{\rm EH} &= -\frac{S_0}{16\pi G_N}\int _{\Sigma _2} \CR \ , \\
    I_{\rm bdy} &= - \frac{S_0}{8\pi G_N}\int_{\partial\Sigma_2} \CK
    - \frac{\Phi_b}{8\pi G_N}\int_{\partial \Sigma_2} \CK + I_{\rm dil\,bdy} \ .
    \end{aligned}
\end{align}
Below we set $4G_N = 1$ for simplicity.
$\Phi$ represents the dilaton and $U(\Phi)$, $V(\Phi)$ are generic functions of $\Phi$. 
The terms proportional to $S_0$ detects the topology of $\Sigma_2$ through the Euler characteristic. 
The JT gravity can be recovered by setting $U(\Phi)$, $V(\Phi)$ to special values:
\begin{align}\label{JT_parametrization}
    {\rm JT\; gravity:}\quad  \Phi = \phi\ ,\quad U = 0\ , \quad V = \Lambda \ .
\end{align}
The cosmological constant $\Lambda$ will be set to $\Lambda = 2$ for AdS$_2$.

We use the replica trick to derive a refined version of the R\'{e}nyi entropy \cite{Dong:2016fnf,Nakaguchi:2016zqi} (also known as the improved R\'enyi/modular entropy):
\begin{align}
\label{eq:defRenyi}
    \Tilde{S}^{(n)} \equiv \partial_\frac{1}{n} \left(\frac{1}{n} \log\, \Tr{\,\rho^n}\right)\ .
\end{align}
In quantum field theory, the calculation of $\Tr{\,\rho^n}$ amounts to computing the partition function on the replica manifold $\widetilde{\CM}_n$ which is the $n$-fold cover of the original spacetime.
We assume the system has the replica $\BZ_n$ symmetry that shifts the $i^{th}$ sheet to the $(i+1)^{th}$ modulo $n$.
Then we can define another manifold $\CM_n = \widetilde{\CM}_n/\BZ_n$ which has the same volume as the original space $\widetilde\CM_1$.
Note that the replica manifold $\widetilde{\CM}_n$ has conical singularities only in regions without gravity while regions coupled with gravity are smooth as dictated by the Einstein equation.
On the other hand, the orbifold $\CM_n$ has conical singularities only in the regions with gravity and the no-gravity regions become smooth.
The on-shell actions on $\widetilde{\CM}_n$ and $\CM_n$ are related by \cite{Lewkowycz:2013nqa,Dong:2013qoa,Dong:2016fnf,Nakaguchi:2016zqi}
\begin{align}
    \label{eq:frmcon}
    \frac{1}{n}\,I_{\rm grav}[\widetilde{\CM}_n]
    = I_{\rm grav}\left[\CM_n\right] + \left(1-\frac{1}{n}\right)\CA^{(n)} \ ,
\end{align}
where $\CA^{(n)}$ is the area term from conical singularities.
In the two-dimensional dilaton gravity theory, the area term can be written by the constant term coming from topology of the orbifold and the value for the dilaton at the conical singularities:
\begin{align}
    \CA^{(n)} = \sum _i \left[S_0 + \Phi^{(n)}(w_i)\right] \ ,
\end{align}
where we denote the $n$ dependence for the dilaton explicitly by $\Phi^{(n)}$.
The positions $w_i$ of the conical singularities are determined by the equation of motion which results from extremizing the action on $\CM_n$ with respect to $w_i$ \cite{Almheiri:2019qdq}:
\begin{align}
\label{eq:QES}
    -\left(1-\frac{1}{n}\right)\partial_{w_i}\Phi^{(n)}(w_i) + \partial_{w_i} \left(\frac{\log Z_{\rm CFT}[\widetilde{\CM}_n]}{n}\right) = 0\ .
\end{align}
Note that this reproduces the quantum extremal surface (QES) condition in $n\to 1$:
\begin{align}
    \begin{aligned}
    \label{eq:QESn1}
    \partial_{w_i}[S_0+\Phi(w_i)+ S_{\rm mat}] = 0\ ,
    \end{aligned}
\end{align}
where $\Phi$ represents the dilaton at $n=1$ and the entanglement entropy for the matter CFT is given by $S_{\rm mat}=\lim_{n\to1}\frac{1}{1-n}\log Z_{\rm CFT}[\widetilde{\CM}_n]$.

Now we turn to the derivation of the refined R\'{e}nyi entropy.
It follows from \eqref{eq:frmcon} that the partition function on the replica manifold $\CM_n$ becomes
\begin{align}
    \begin{aligned}
    \label{eq:reppar}
    -\frac{1}{n}\log \Tr{\,\rho^n} 
    &= \frac{1}{n} I_{\rm grav}[\widetilde{\CM}_n] - \dfrac{1}{n} \log Z_{\rm CFT}[\widetilde{\CM}_n]\\
    &= I_{\rm grav}[\CM_n] + \left(1-\frac{1}{n}\right)\CA^{(n)}-\frac{1}{n}\log Z_{\rm CFT}[\widetilde{\CM}_n]\ .
    \end{aligned}
\end{align}
Substituting it into \eqref{eq:defRenyi} we end up with
\begin{align}
    \Tilde{S}^{(n)} 
        &= 
        \CA^{(n)} +
            \frac{1}{n}\frac{\delta \log Z[\widetilde{\CM}_n]}{\delta g_{\mu \nu}} \partial_\frac{1}{n}g_{\mu \nu}
            -\frac{1}{n}\frac{\delta I_{\rm grav}[\widetilde\CM_n]}{\delta \Phi}\,\partial_\frac{1}{n}\Phi + \partial_\frac{1}{n}\left(\frac{1}{n}\log Z_{\rm CFT}[\widetilde{\CM}_n]\right)\bigg|_{g} \notag\\
        &= 
        \sum _i \left[S_0 + \Phi^{(n)}(w_i)\right] + \Tilde{S}_{\rm mat}^{(n)} \ , \label{eq:Renyi}
\end{align}
where the total action is $\log Z[\widetilde{\CM}_n]=-I_{\rm grav}[\widetilde\CM_n]+\log Z_{\rm CFT}[\widetilde{\CM}_n]$ and we have imposed the equations of motion on $\widetilde\CM_n$ in the second line.
The refined R\'{e}nyi entropy of the matter
\begin{align}
    \Tilde{S}_{\rm mat}^{(n)}=\partial_\frac{1}{n}\left(\frac{1}{n}\log Z_{\rm CFT}[\widetilde{\CM}_n]\right)\bigg|_{g}\ ,
\end{align}
is calculated by taking the $n$-derivative of the matter partition function on the replica manifold $\widetilde{\CM}_n$ with the metric determined by the equation of motion.

Having the application of the formula to the Hawking radiation in mind 
we consider the JT gravity on an AdS$_2$ black hole and couple it to an exterior flat space without gravity that plays a role of a heat bath.
We also impose the transparent boundary condition for the matter CFT at the boundary between the gravity region and the flat bath.
The time evolution of the black hole is probed through the entanglement (R\'enyi) entropy of a radiation region \text{Rad} in the bath, where we collect the radiation particle from the black hole.
The partition function for the matter on the replica geometry $\widetilde{\CM}_n$ reduces to the correlation function of twist operators at the branch points (conical singularities) of the replica $\BZ_n$ symmetry on the original sheet $\widetilde{\CM}_1$.
If we make gravity non-dynamical, all the branch points are located on the boundaries of the radiation region.
With gravity turned on in a part of the spacetime, however, there appear the $\BZ_n$-fixed points in the gravity region which form additional branch points at the boundaries of the so-called island region $\text{I}$.
This implies that, after the island formation, the matter entropy should be evaluated on the union of the island region in the gravitational region and the radiation region in the flat bath.
In the gravitational path integral derivation, it is manifest that the replica geometry becomes a replica wormhole constructed by gluing $n$ copies of $\widetilde\CM_1$ along the island region cyclically.
In particular \eqref{eq:Renyi} amounts to the island formula in the $n = 1$ limit:
\begin{align}\label{island_formula}
    S(\text{Rad}) =\underset{\text{I}}{\min}\left[
    \underset{\text{I}}{\mathrm{ext}}\left[
    \sum_{\partial \text{I}}\left(S_0 + \Phi(\partial \text{I})\right)
    + S_\text{mat}(\text{Rad} \cup \text{I})
    \right]\right]\ .
\end{align}
We also impose the QES condition \eqref{eq:QESn1} as the extremization over the island region.
The replica partition function is evaluated in the path integral by summing up all replica geometries, but in semiclassical approximation the on-shell solution with least action becomes dominant, justifying the minimal prescription above.
Thus the island formula gives a complete answer for a gravitational fine-grained (entanglement) entropy in the semiclassical limit.

Back to the discussion on the refined R\'{e}nyi entropy, we should also impose the extremization condition \eqref{eq:QES} and evaluate the formula \eqref{eq:Renyi} on the dominant saddle solution in the gravitational path integral in the same way as the island formula for the entanglement entropy.
The simple-looking formula \eqref{eq:Renyi} is however not more pragmatic in use than the island formula for at least two reasons.
One has to solve the dilaton equation of motion and fix the positions $w_i$ of the conical singularities by the condition $\eqref{eq:QES}$ at the same time.
Namely, we need a solution of the dilaton in the presence of conical singularities located at generic points.
Moreover, if the gravity region is coupled to a non-gravitating bath we also have to solve the conformal welding problem to calculate the matter partition function.
We will revisit the latter issue in more detail in section \ref{ss:welding}.

\subsection{Capacity of entanglement}\label{ss:capacity_ent}
Since we have already derived the refined R\'{e}nyi entropy \eqref{eq:Renyi} in the previous subsection,
it is straightforward to obtain the $n^{th}$ capacity in the dilaton gravity coupled to a matter CFT.
The $n^{th}$ capacity of entanglement is defined as the derivative with respect to $n$ \cite{Yao:2010woi,Nakaguchi:2016zqi}:
\begin{align}\label{capacity_def}
    C^{(n)} \equiv -n\,\frac{\partial \tilde{S}^{(n)}}{\partial n}\ .
\end{align}
Note that $\tilde{S}^{(n)}$ is evaluated on a gravitational saddle satisfying the equation \eqref{eq:QES}.
For the capacity \eqref{capacity_def} to be well-defined as the derivative of $\tilde{S}^{(n)}$, we assume that the gravitational saddle varies smoothly for an infinitesimal change of $n$.

The contribution from the dilaton gravity can be written as
\begin{align}\label{eq:nth_capacity}
    \frac{\partial \CA^{(n)}}{\partial n}
        = \sum_i \left(\frac{\partial \Phi^{(n)}(w)}{\partial n}\bigg|_{w=w_i}  + \sum_j\,\frac{\partial \Phi^{(n)}(w_i)}{\partial w_j}\frac{\partial w_j}{\partial n}\right)\ .
\end{align}
The first and second terms originate from the $n$ dependence of the dilaton field and the positions of conical singularities respectively.
Including the contribution from the matter CFT, the capacity becomes
\begin{align}
    C^{(n)} 
        =-n\sum_i \frac{\partial \Phi^{(n)}(w)}{\partial n}\bigg|_{w=w_i}  + C_{\rm mat}^{(n)} - n\sum_i \partial_{w_i}\left(\Phi^{(n)}(w_i)+ \tilde{S}^{(n)}_{\rm mat}\right)\frac{\partial w_i}{\partial n} \ ,
\end{align}
where $C_{\rm mat}^{(n)} \equiv -n\,{\partial \tilde{S}_{\rm mat}^{(n)}}/{\partial n}$ is the $n^{th}$ capacity for the matter part.

In what follows we will be interested in $C\equiv C^{(n=1)}$, which we call the capacity of entanglement for simplicity, that simplifies due to the QES condition \eqref{eq:QESn1}:
\begin{align}\label{eq:generalcap}
    C 
    =  -\sum_i \frac{\partial \Phi^{(n)}(w)}{\partial n}\bigg|_{n=1,w=w_i} + C_{\rm mat}\ .
\end{align}
It follows from our assumption stated below \eqref{capacity_def} that the gravitational saddle the dilaton (and the matter capacity) depends on varies smoothly with respect to the replica parameter around $n=1$. Thus the capacity \eqref{eq:generalcap} should be evaluated on the same saddle as the generalized entropy determined by the island formula \eqref{island_formula}.

A few comments are in order on the formula \eqref{eq:generalcap}:
\begin{itemize}
    \item 
    While the island formula can determine a dominant saddle solution in the gravitational path integral the capacity \eqref{eq:generalcap} does not play a similar role, but rather be applied to the saddle chosen by the island formula.
    \item
    Since we imposed the QES condition \eqref{eq:QESn1} for $n=1$ to derive \eqref{eq:generalcap} from \eqref{eq:nth_capacity} 
    the positions $w_i$ of the conical singularities in the formula \eqref{eq:generalcap} are to be fixed by \eqref{eq:QESn1} as well.
    \item 
    The first term in \eqref{eq:generalcap} represents the $n$-derivatives of the dilaton at the conical singularities, which requires us to solve the dilaton equation of motion in the replica geometry at least of order $O(n-1)$.
    \item 
    In contrast to the island formula, the capacity has no topological term $S_0$, which essentially makes it jump around the Page time.
    We will explain this mechanism in section \ref{sec:disc of CoE}.
    \item
    The conformal welding problem complicates the calculation of $C_\text{mat}$ in \eqref{eq:generalcap} as will be touched on in section \ref{ss:welding}. 
\end{itemize}

\subsection{Discontinuity of capacity due to topology change}\label{sec:disc of CoE}

The island formula \eqref{island_formula} picks up a dominant saddle with the least entropy, so if two saddle solutions compete with each other they have the same entropy at the Page time.
For an evaporating black hole, we expect the black hole solution is dominant at early time, but it is taken over by the replica wormhole solution at late time.
The black hole entropy $S_\text{BH}$ grows linearly in time while the generalized entropy $S_\text{RWH}$ of the replica wormhole typically saturates or decreases, reproducing the expected Page curve.
Incidentally the entropy is always continuous at the Page time where the two entropies coincide: $S_\text{BH} = S_\text{RWH}$.

In the previous section \ref{ss:capacity_ent} we derived the formula \eqref{eq:generalcap} for calculating the (first) capacity of entanglement in the two-dimensional dilaton gravity.
In contrary to the island formula for the entropy, \eqref{eq:generalcap} does not fix the dominant saddle for a radiation region of interest by itself.
The capacity is rather given by evaluating \eqref{eq:generalcap} on a dominant saddle chosen by the island formula.
Hence there is no reason for the capacity to be continuous at the Page time across which the black hole phase is superseded by the replica wormhole phase.

Let us consider the case where conformal matter is given by free fermions and there are black hole and replica wormhole phases with the entropies:
\begin{align}
    \begin{aligned}
        S_\text{BH} &= S_\text{ferm}(\text{Rad}) \ ,\\
        S_\text{RWH} &= \sum_{w_i \in \partial \text{I}} \left[S_0 + \Phi(w_i)\right] + S_\text{ferm}(\text{Rad}\cup \text{I}) \ .
    \end{aligned}
\end{align}
It follows from \eqref{eq:generalcap} that the capacity of each phase takes the following form:
\begin{align}
    \begin{aligned}
        C_\text{BH} &= C_\text{ferm}(\text{Rad}) \ ,\\
        C_\text{RWH} &= - \sum_{w_i \in \partial \text{I}} \partial_n \Phi^{(n)}(w_i)|_{n=1} + C_\text{ferm}(\text{Rad}\cup \text{I}) \ .
    \end{aligned}
\end{align} 
Note that the topological term $S_0$ is invisible to the capacity $C_\text{RWH}$ itself as opposite to the entropy $S_\text{RWH}$.
The R\'enyi entropy of $c$ Dirac fermions for $p$ intervals $\bigcup_{i=1}^p [u_i, v_i]$ on a conformally flat space with the metric $\d s^2 = \Omega^{-2}\,\d w\,\d \bar w$ can be read off from the result on the flat space \cite{Casini:2005rm} as
\begin{align}\label{eq:matter Renyi entropy}
    S^{(n)}_\text{ferm} = \frac{c}{6}\left(1 + \frac{1}{n}\right)\,\log \left[ \frac{\prod_{i,j} |u_i - v_j| }{\varepsilon^p\, \prod_i \Omega(u_i)^\frac{1}{2}\,\Omega(v_i)^\frac{1}{2}\, \prod_{i< j} |u_i - u_j|\,|v_i - v_j|}\right] \ .
\end{align}
The arguments of the logarithm can depend on the replica parameter $n$ due to the conformal welding map as we will review in section \ref{ss:welding}. 
This is a highly non-trivial problem to solve even in the vicinity of $n=1$, so we assume for a moment there are cases where the conformal welding problem becomes trivial and there still exists a phase transition between the black hole and replica wormhole.
In such a case, the $n$ dependence simplifies in the R\'enyi entropy and the capacity turns out to equal to the entanglement entropy \cite{deBoer:2018mzv}:
\begin{align}
     C_\text{ferm} = S_\text{ferm} \qquad (\text{when the welding problem is trivial}) \ .
\end{align}
It follows that the discontinuity $\Delta C\equiv C_\text{BH} - C_\text{RWH}$ of the capacity at the Page time ($S_\text{BH} = S_\text{RWH}$) becomes
\begin{align} \label{eq:capdiscont}
    \Delta C = \sum_{w_i \in \partial I} \left[ S_0 + \Phi(w_i) + \partial_n \Phi^{(n)}(w_i)|_{n=1}\right] \ .
\end{align}
Since $\Phi^{(n)}(w_i)$ and $\partial_n \Phi^{(n)}(w_i)|_{n=1}$ are independent of $S_0$, the discontinuity $\Delta C$ can be made nonzero for some choice of $S_0$. 
The topological term $S_0$ is generated by topology change between the black hole and replica wormhole. It implies that the topology change can be an essential source of the discontinuous jump of the capacity at the Page time.

While we could not find an illustrating example where all the assumptions made so far hold,
we expect that the $n$ dependence of the conformal factor $\Omega$ does not change the above consideration drastically and the capacity is discontinuous in general.

\subsection{Conformal welding problem}\label{ss:welding}
Behind the simplicity of the formula for the capacity of entanglement \eqref{eq:generalcap}, it involves practical difficulties coming from the conformal welding problem. 
Here we give a brief review of the conformal welding problem and how it complicates the calculation of the capacity of entanglement in the JT gravity on the AdS${}_2$ spacetime coupled to a non-gravitating thermal bath \cite{Sharon2006,Almheiri:2019qdq,Goto:2020wnk}.

\begin{figure}[ht!]
	\centering
	\begin{tikzpicture}[transform shape]
        \begin{scope}[xshift=-3.5cm,yshift=0cm]
            \draw[black!100, thick] (-2.5,-2.5) rectangle (2.5,2.5);
            \filldraw[cyan!30, opacity=0.4] (0,0) circle [radius=1.4]; 
            \draw[black!100, very thick,densely dashdotted] (0,0) circle [radius=1.4] node[thick, above] {$|w|< 1$};
            \draw[->,->=stealth, black!100,above] (-0.5,-0.7) node[right,thick, black!100,font=\small] {$w=e^{\i\theta (\tau)}$} to[out=180,in=70] (-0.98,-0.95); 
            \draw[->,->=stealth, black!100,above] (-0.5,-1.7) node[right,thick, black!100,font=\small] {$v=e^{\i\tau}$} to[out=180,in=-90] (-0.8,-1.2); 
            \node[black!100,font=\small] at (-1.7,2) {$|v|>1$}; 
        \end{scope}

        \begin{scope}[xshift=3.5cm,yshift=0cm]
            \draw[black!100, thick] (-2.5,-2.5) rectangle (2.5,2.5);
            \filldraw[cyan!30, opacity=0.4,rounded corners] (-1.5,0.2) to[out=45,in=-200] (1,1) to[out=-70,in=95] (0.7,-0.3) to[out=-85,in=120] (1.2,-1.3) to[out=-60,in=-20] (-0.2,-1.6) to[out=160,in=-135] (-1.5,0.2);
            \draw[black!100, very thick,densely dashdotted, rounded corners] (-1.5,0.2) to[out=45,in=-200] (1,1) to[out=-70,in=95] (0.7,-0.3) to[out=-85,in=120] (1.2,-1.3) to[out=-60,in=-20] (-0.2,-1.6) to[out=160,in=-135] (-1.5,0.2);
            \draw[black!100, thick] (2.4,2) -- (2,2) node[above right]{$z$} -- (2,2.4) ; 
        \end{scope}

        \draw[->,->=stealth, very thick] (-2.8,0) to[out=-15,in=-165] node[midway,above, black!100, thick] {$G(w)$} (2.8,0); 
        \draw[->,->=stealth, 
        very thick] (-2.8,1.7) to[out=10,in=170] node[midway,above, black!100, thick] {$F(v)$} (2.8,1.7); 
  
	\end{tikzpicture}
	\caption{In the conformal welding problem two regions parametrized by $|w|\leq1$ and $|v|\geq1$ are glued together along their boundaries, by finding holomorphic functions: $G(w),\, F(v)$ 	subject to the boundary condition \eqref{eq:welding maps}. In the JT gravity on an AdS${}_2$ black hole coupled to a flat bath region, we parametrize the former by $w$ and the latter by $v$.}
	\label{fig:conformal welding problem}
\end{figure}
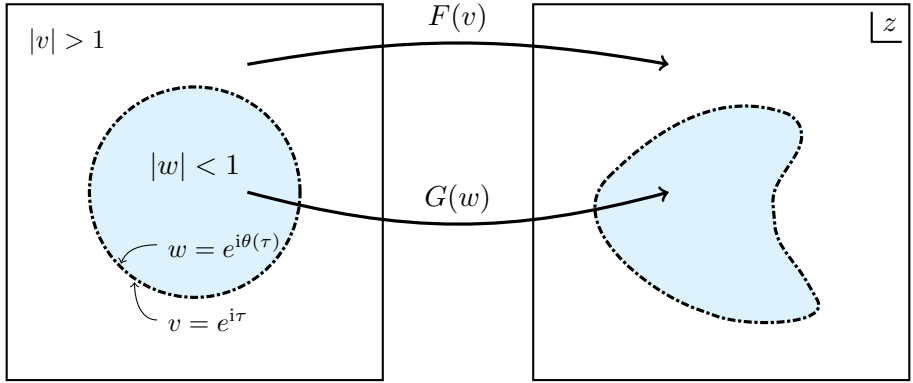

The conformal welding problem is to create a new Riemann surface out of two Riemann surfaces by cutting a circle out of each and gluing them smoothly at their boundary circles.
This problem is phrased in the language of complex analysis, by parametrizing these two Riemann surfaces, each with a disk removed, by $|w|\leq 1$ and $|v|\geq1$, so that $|w|=1 $ and $|v|=1$ represent their respective boundary circles.
One cannot naively extend one coordinate holomorphically to the other beyond its boundary.
Nevertheless, the Riemann mapping theorem assures the existence of holomorphic maps $G(w)$ and $F(v)$ that can map both regions $|w|\leq 1$ and $|v|\geq 1$ onto a complex $z$-plane with the images of the both boundaries glued together (see figure \ref{fig:conformal welding problem}):\footnote{There is an $\mathrm{SL}(2,\mathbb{R})$ redundancy for each map.}
\begin{align}\label{eq:welding maps}
    \begin{aligned}
    z=\begin{dcases}
    G(w)  & \text{for } |w|\leq 1\\
    F(v) & \text{for } |v|\geq 1
    \end{dcases}\ ,\qquad \text{with}\quad G\left(e^{\i\theta(\tau)}\right)=F\left(e^{\i\tau}\right) \quad \text{at the boundaries} \ .
    \end{aligned}
\end{align}
The whole new Riemann surface is covered by the $z$ coordinates. In general, it is impossible to find analytic expressions of the welding maps: $F$ and $G$, since they depend non-locally on the boundary mode $\theta(\tau)$.

Let us turn to a more concrete setup where the JT gravity on an AdS${}_2$ spacetime is coupled to a flat bath region without gravity.
The conformal welding problem is trivial as the two regions can be glued smoothly just by imposing transparent boundary condition.
However, it becomes complicated when it comes to the evaluation of entanglement entropy and the capacity of entanglement.
As discussed around \eqref{island_formula} the calculation of the entanglement entropy of the Hawking radiation via the replica method amounts to inserting twist operators on the bath region and additional conical singularities induced dynamically due to non-perturbative effects in the gravitating region.
The gluing procedure is quite non-trivial in the presence of conical singularities that distort the geometry of the gravitating region such that its boundary mode depends on $n$ through the equation of motion from the transparent boundary condition.

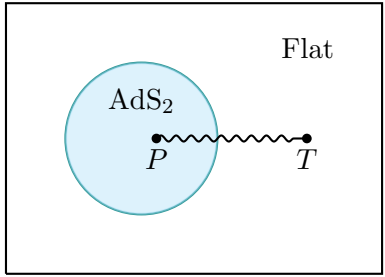
\begin{figure}
\begin{center}
    \begin{tikzpicture}
        \draw[thick, black!100] 
        (-2,-1.8) -- (3,-1.8) -- (3,1.8) -- (-2,1.8) -- (-2,-1.8);
        \draw[thick,teal!80](-0.2,0) circle [radius=1.01];
        \filldraw[cyan!30, opacity=0.4,thick] (-0.2,0) circle [radius=1];
        \draw[thick, decorate,decoration={snake,amplitude=.4mm,segment length=2mm,post length=1mm}] node [below,font = \normalsize]{$P$} (0,0) -- (2,0) node [below, font=\normalsize]{$T$};
        \draw (2.5,1.2) node[left,font = \normalsize]{Flat};
        \draw (-0.2,0.5) node[font=\normalsize]{AdS$_2$};
        \filldraw[thick] (0,0) circle (0.05);
        \filldraw[thick] (2,0) circle (0.05);
    \end{tikzpicture}
\end{center}
\caption{This figure shows the single interval $[P,T]$ in the JT gravity on AdS${}_2$ plus a non-gravitating bath. As we will see in section \ref{ss:entropy}, we can interpret this as the Euclidean island/replica wormhole configuration of our main interest throughout this paper.}
\label{fig:euclideanset}
\end{figure}

To illustrate the difficulty associated with the conformal welding problem more concretely, let us focus on a single interval $[P,T]$ stretching from a point $P$ inside the $\mathrm{AdS}{}_2$ region $|w|\leq1$ to a point $T$ in the flat bath region $|v|\geq1$ as in figure \ref{fig:euclideanset}.
To evaluate the capacity of entanglement \eqref{eq:generalcap} on the interval, it is necessary to obtain the dilaton solution at $P$ and the $n$ dependent conformal welding maps up to the first order in $(n-1)$.
The latter $n$ dependence comes in the matter R\'enyi entropy as
\begin{align}\label{eq:Renyi welding}
    S^{(n)}_{\text{mat}}=\frac{c}{12}\left(1+\frac{1}{n}\right)\,\log \left[\frac{(F(B)-G(A))^2}{\varepsilon^2\,F'(B)\, G'(A)\,\Omega_{\text{ads}}(A)}\right]\ ,
\end{align}
where we parametrized $P$ by $w=A$ and $T$ by $v$ in their respective coordinate systems.
$\Omega_{\mathrm{ads}}(A)$ is an $n$ independent Weyl factor coming from the $\mathrm{AdS}{}_2$ region.
The entanglement entropy $S_\text{mat}= S^{(n=1)}_{\text{mat}}$ is independent of the welding map as they are trivial, $F=G=1$ for $n=1$ \cite{Almheiri:2019qdq}, but the capacity of entanglement $C^{(n=1)}_\text{mat}=-2\partial_n S_\text{mat}^{(n)}\big|_{n=1}$ depends on $n$ through the derivatives $\left.\partial_n F(B)\right|_{n=1},\left.\partial_n G(A)\right|_{n=1}$.
So the calculation of the capacity of entanglement requires us to solve the conformal welding problem and to find the explicit form of the welding maps at least of order $O(n-1)$.

Even for the simplest single interval case, it is a rather hard task to solve the conformal welding problem analytically once the replica parameter $n$ is shifted away from one, except for the high temperature limit $\kappa\propto \beta\to 0$ where the welding maps turn out to be trivial 
(see appendix \ref{ss:fromreplica}).\footnote{The numerical analysis of the R\'enyi entropy for $1\le n <2$ is performed in \cite{Mirbabayi:2020fyk}, from which the capacity of entanglement for a single interval can be read off numerically even at finite temperature. We thank M.\,Mirbabayi for the correspondence on this point.} 
In section \ref{sec:high temp limit AdS2}, we will implement the formula \eqref{eq:generalcap} and calculate the capacity of entanglement for a single interval in the high temperature limit. 
We will also reproduce the same result by another method using the boundary mode in appendix \ref{ss:fromreplica}.

\section{Example: AdS$_2$ eternal black holes at high temperature}\label{sec:high temp limit AdS2}

We have derived the formula \eqref{eq:generalcap} for the capacity of entanglement in the two-dimensional dilaton gravity through the careful analysis of the gravitational path integral on the replica geometry.
Despite the practical difficulties due to the conformal welding problem, the capacity of entanglement is expected to exhibit a discontinuity around the Page time and serve as a new probe of the island formation/replica wormhole. 

In this section, we consider an analytically tractable model of a radiating black hole to which the capacity formula \eqref{eq:generalcap} can be applied: the high temperature limit of the AdS$_2$ eternal black hole coupled to a flat bath in the JT gravity.
In section \ref{ss:LorentzianAdS2} we start with reviewing the Lorentzian geometry of the model.
Then we proceed to study the entanglement (R\'enyi) entropy and the capacity of entropy in section \ref{ss:entropy} and \ref{ss:capacity_ads2} respectively.

While the replica wormhole saddle always dominates and there is no phase transition to the black hole saddle in this example 
the capacity has an alternative way to calculate using the boundary mode as performed in appendix \ref{ss:fromreplica}, so it serves as a consistency check of the formula \eqref{eq:generalcap}.
Finally, in section \ref{ss:thermal} we compare the entropy and capacity of entanglement with the thermal counterparts and find they match at high temperature.

\subsection{Coordinate system}\label{ss:LorentzianAdS2}
We introduce the coordinate $y_R^\pm$ and $y_L^\pm$ in the right/left Rindler wedges \cite{Almheiri:2019yqk}:
\begin{align}
    \begin{aligned}
    y_R^\pm = t_R \pm \sigma_R\ ,\qquad
    y_L^\pm = t_L \mp \sigma_L\ .
    \end{aligned}
\end{align}
In these coordinates, the classical solution of the JT gravity on the AdS$_2$ eternal black hole with inverse temperature $\beta$ is given in the right Rindler wedge  by
\begin{align}\label{eq:Lorentzian AdS2 BH}
    \d s_{\rm ads}^2 = -\left(\frac{2\pi}{\beta}\right)^2 \frac{\d y_R^+\, \d y_R^-}{\sinh^2\frac{\pi(y_R^--y_R^+)}{\beta}}\ ,\qquad
    \phi = \frac{2\pi}{\beta}\frac{\phi_r}{\tanh \frac{\pi(y_R^--y_R^+)}{\beta}}\ ,
\end{align}
and similarly for the left Rindler wedge by replacing $y_R^\pm \to y_L^\pm$, where we use the notation for the JT gravity \eqref{JT_parametrization}.
The thermal baths attached to both sides of the AdS$_2$ boundaries are flat spaces.
We introduce the cutoff $\sigma_R = -\epsilon$ for the right conformal boundary of the AdS$_2$.
The metrics outside the AdS$_2$ spacetime are fixed so that the boundaries are glued smoothly.
The metric outside the right Rindler wedge is
\begin{align}
    \d s_{\rm bath}^2 = -\frac{\d y_R^+\, \d y_R^-}{\epsilon^2}\ ,
\end{align}
and similarly for the left side.

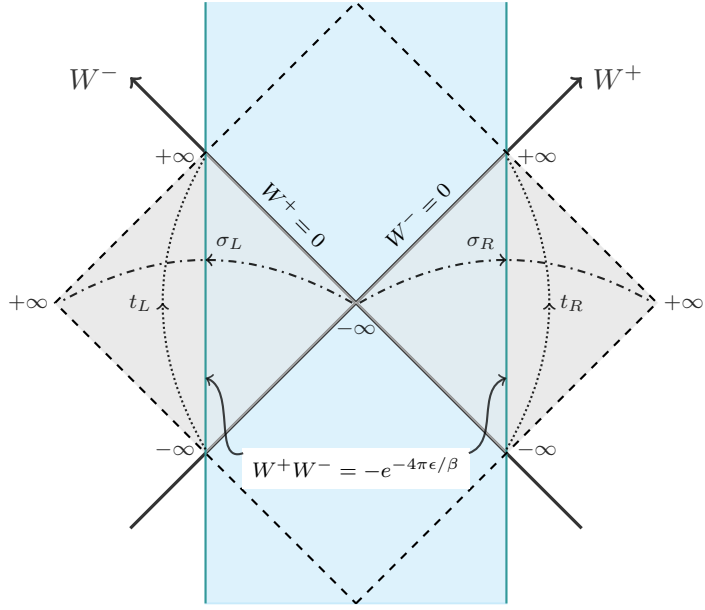
\begin{figure}
    \centering
    \begin{tikzpicture}[transform shape]

        \filldraw[cyan!30, thick,opacity=0.4] (-2,6) -- (-2,-2) -- (2,-2) -- (2,6);
        \filldraw[gray!30, thick,opacity=0.4] (2,0) -- (4,2) -- (2,4) -- (2,0);
        \filldraw[gray!30, thick,opacity=0.4] (-2,0) -- (-4,2) -- (-2,4) -- (-2,0);
        \draw[very thick,black!80,->] (-3,-1) -- (-1,1)  --node [black!100,midway,above, rotate=45, font=\scriptsize] {$W^-=0$} (3,5) node [right, font=\normalsize] {$W^+$} ;
        \draw[very thick,black!80,->] (3,-1) -- (1,1) -- node [black!100,midway,above, rotate=-45, font=\scriptsize] {$W^+=0$}  (-3,5) node [left, font=\normalsize] {$W^-$} ;
        \filldraw[gray!20, opacity=0.5] (-2,0)--(0,2)--(-2,4)--(-4,2);
        \filldraw[gray!20, opacity=0.5] (2,0)--(0,2)--(2,4)--(4,2);
        \draw[teal!80,thick] (-2,-2) --   (-2,6);
        \draw[teal!80,thick] (2,-2) -- (2,6);
        \draw[thick, dashed, opacity=1] (0,-2)  -- (2,0) --  (4,2) -- (2,4)   -- (0,6) -- (-2,4)  --  (-4,2)  -- ( -2,0)   --  (0,-2);
        \draw[middlearrow={>},black!100, thick,opacity=0.85,densely dotted] (2,0.05) to[out=60,in=-60] node[midway, right,font=\scriptsize] {$t_R$} (2,3.95);
        \node[black!100, thick, right,font=\scriptsize,opacity=0.85]  at (2,0.05) {$-\infty$};
        \node[black!100, thick, right,font=\scriptsize,opacity=0.85]  at (2,3.95) {$+\infty$};
        \draw[middlearrow={>},black!100, thick,opacity=0.85,dashdotted] (0.05,2) to[out=30,in=150] node[midway, above left,font=\scriptsize] {$\sigma_R$} (3.95,2);
        \node[black!100, thick, right ,font=\scriptsize,opacity=0.85]  at (3.95,2) {$+\infty$};
        \node[black!100, thick,opacity=0.85, below ,font=\scriptsize]  at (0,1.9) {$-\infty$};

        \draw[middlearrow={>},black!100, thick,opacity=0.85,densely dotted] (-2,0.05) to[out=120,in=-120] node[midway, left,font=\scriptsize] {$t_L$} (-2,3.95);
        \node[black!100, thick,opacity=0.85, left,font=\scriptsize]  at (-2,0.05) {$-\infty$};
        \node[black!100, thick,opacity=0.85, left,font=\scriptsize]  at (-2,3.95) {$+\infty$};

        \draw[middlearrow={>},black!100, thick,opacity=0.85,dashdotted] (-0.05,2) to[out=150,in=30] node[midway, above right,font=\scriptsize,black!100, thick,opacity=0.85] {$\sigma_L$} (-3.95,2);
        \node[left ,font=\scriptsize,black!100, thick,opacity=0.85]  at (-3.95,2) {$+\infty$};

    	\node[black!100,very thick, font=\scriptsize,rectangle,fill=white!100] at (0,-0.2) {$W^+W^-= -e^{-4\pi\epsilon/\beta}$};
     	\draw[black!100,->,->=stealth,thick,opacity=0.75] (1.5,0) to[out=20,in=200] (2,1) ;
 	 	\draw[black!100,->,->=stealth,thick,opacity=0.75] (-1.5,0) to[out=200,in=-20] (-2,1) ;             
    \end{tikzpicture}
	\caption{The AdS$_2$ eternal black hole in Lorentzian signature (colored in blue). We introduce three coordinates $y_R^{\pm}$, $y_L^{\pm}$ and $W^{\pm}$. We show the patch described by $y_R^{\pm}$, $y_L^{\pm}$ as the shaded regions. $W^{\pm}$ covers inside the dashed square.}
	\label{fig:lorentzAdS}
\end{figure}

To cover the full Cauchy slice of the AdS$_2$ plus thermal bath spacetime let us introduce new coordinates $W^\pm$ as follows:
\begin{align}
    W^\pm = \pm e^{\pm \frac{2\pi}{\beta} y_R^\pm}\ ,\qquad
    W^\pm = \mp e^{\mp \frac{2\pi}{\beta} y_L^\pm}\ .
\end{align}
The right and left AdS$_2$ boundaries are characterized by $W^\pm = \pm e^{\pm \frac{2\pi}{\beta}(t_R\mp \epsilon)}$ and $W^\pm = \mp e^{\mp \frac{2\pi}{\beta}(t_L\pm \epsilon)}$ respectively or more simply by the condition $W^+W^- = - e^{-\frac{4\pi\epsilon}{\beta}}$ (see figure \ref{fig:lorentzAdS}).
The new coordinates cover the whole spacetime with the conformally flat metric
\begin{align}
    \begin{aligned}
    \d s^2 = - \Omega^{-2}\, \d W^+ \d W^- \ ,
    \end{aligned}
\end{align}
where the conformal factor $\Omega$ is given in the AdS and flat bath regions respectively by
\begin{align}\label{ConformalFactors}
    \Omega_\text{ads} = \frac{1 + W^+W^-}{2}\ , \qquad \Omega_\text{bath} = \frac{2\pi \epsilon}{\beta}\,\sqrt{-W^+W^-} \ .
\end{align}

\subsection{Entanglement entropy}
\label{ss:entropy}

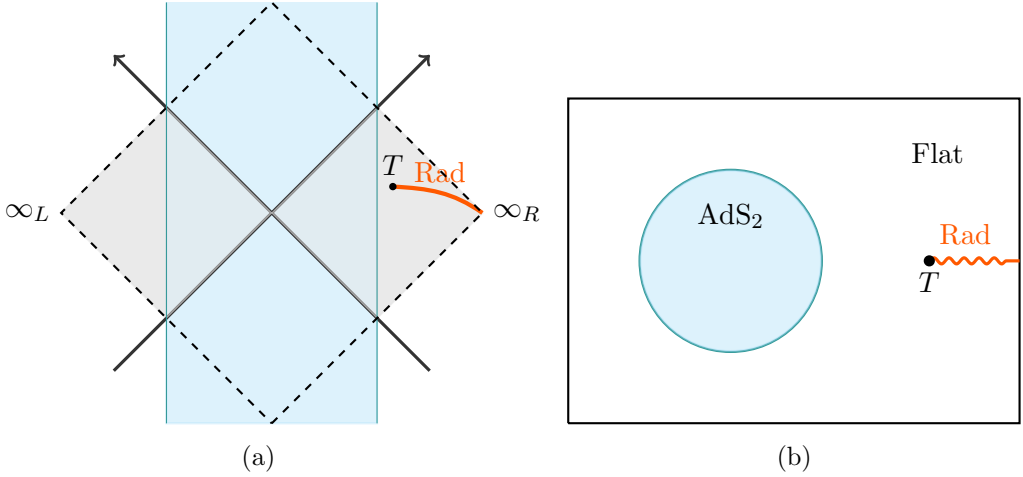
\begin{figure}[htbp]
  \begin{minipage}[b]{0.45\linewidth}
    \centering
    \begin{tikzpicture}[scale=0.7]
        \filldraw[cyan!30, thick,opacity=0.4] (-2,6) -- (-2,-2) -- (2,-2) -- (2,6);
        \filldraw[gray!30, thick,opacity=0.4] (2,0) -- (4,2) -- (2,4) -- (2,0);
        \filldraw[gray!30, thick,opacity=0.4] (-2,0) -- (-4,2) -- (-2,4) -- (-2,0);
        \draw[very thick,black!80,->] (-3,-1) -- (-1,1)  --   (3,5);
        \draw[very thick,black!80,->] (3,-1) -- (1,1) --   (-3,5);
        \filldraw[gray!20, opacity=0.5
        ] (-2,0)--(0,2)--(-2,4)--(-4,2);
        \filldraw[gray!20, opacity=0.5
        ] (2,0)--(0,2)--(2,4)--(4,2);
  
        \draw[teal!80] (-2,-2) --   (-2,6);
        \draw[teal!80] (2,-2) --  (2,6);
        \draw[thick](-4,2) node [thick,left] {$\infty_L$};
        \draw[thick](4,2) node [thick,right] {$\infty_R$};
       
        \draw[orange!70!red, ultra thick,rounded corners] (2.3,2.5) node[above,black!100] {$T$} to[out=-3,in=145] node[midway, above,thick] {$\text{Rad}$} (4,2); 
        
        \filldraw[thick] (2.3,2.5) circle (0.05);
        \draw[thick, dashed, opacity=1] (0,-2)  -- (2,0) --  (4,2) -- (2,4)   -- (0,6) -- (-2,4)  --  (-4,2)  -- ( -2,0)   --  (0,-2);
    \end{tikzpicture}
    \subcaption{}
    \label{fig:Rad_Lorentz}
  \end{minipage}
  \begin{minipage}[b]{0.45\linewidth}
    \centering
    \begin{tikzpicture}[scale=1.2]
        \draw[thick,black!100] (-2,-1.8) -- (3,-1.8) -- (3,1.8) -- (-2,1.8) -- (-2,-1.8);
        \draw[thick,teal!80](-0.2,0) circle [radius=1.01];
        \filldraw[cyan!30, opacity=0.4,thick] (-0.2,0) circle [radius=1];
    
        \draw[orange!70!red, very thick,  decorate,decoration={snake,amplitude=.4mm,segment length=2mm,post length=1mm}]
        (2,0) node [below, black, font=\normalsize]{$T$} -- (3,0);
        \draw (2,0.3) node[orange!70!red, font=\normalsize,below,right]{\text{Rad}};
        \draw (2.5,1.2) node[left,font = \normalsize]{Flat};
        \draw (-0.2,0.5) node[font=\normalsize]{AdS$_2$};
        \filldraw[thick] (2,0) circle (0.05);
    \end{tikzpicture}
    \subcaption{}
    \label{fig:Rad_Euclid}
  \end{minipage}
  \caption{(a) The Lorentzian setup of a semi-infinite line region $\text{Rad}=[T,\infty_R)$ in the flat bath. The entanglement entropy of \text{Rad} is IR divergent. (b) The Euclidean configuration where AdS$_2$ black hole (hyperbolic disk) is glued to the flat space.
  }\label{fig:Rad}
\end{figure}

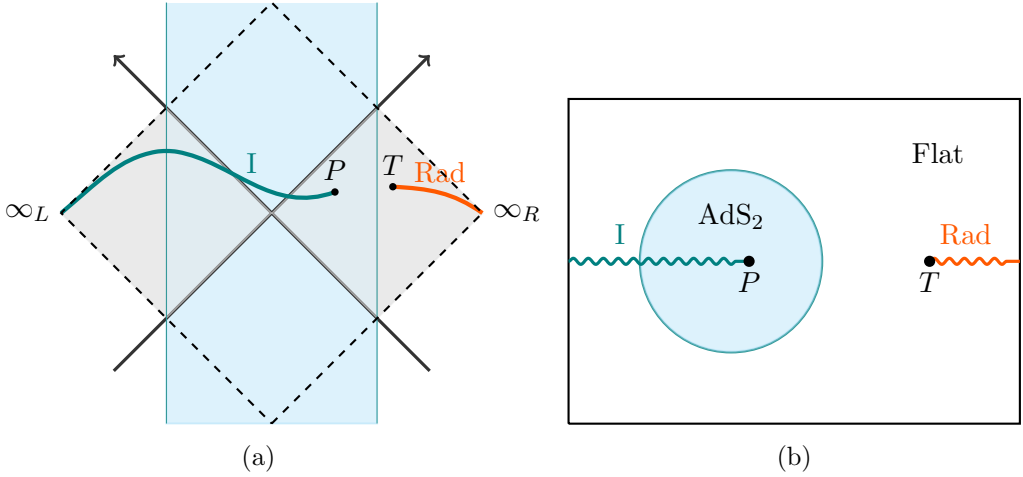
\begin{figure}[htbp]
  \begin{minipage}[b]{0.45\linewidth}
    \centering
    \begin{tikzpicture}[scale=0.7]
        \filldraw[cyan!30, thick,opacity=0.4] (-2,6) -- (-2,-2) -- (2,-2) -- (2,6);
        \filldraw[gray!30, thick,opacity=0.4] (2,0) -- (4,2) -- (2,4) -- (2,0);
        \filldraw[gray!30, thick,opacity=0.4] (-2,0) -- (-4,2) -- (-2,4) -- (-2,0);
        \draw[very thick,black!80,->] (-3,-1) -- (-1,1)  --(3,5);
        \draw[very thick,black!80,->] (3,-1) -- (1,1) -- (-3,5);
        \filldraw[gray!20, opacity=0.5
        ] (-2,0)--(0,2)--(-2,4)--(-4,2);
        \filldraw[gray!20, opacity=0.5
        ] (2,0)--(0,2)--(2,4)--(4,2);
       
        \draw[teal!80] (-2,-2) --   (-2,6);
        \draw[teal!80] (2,-2) --  (2,6);
        \draw[thick](-4,2) node [thick,left] {$\infty_L$};
        \draw[thick](4,2) node [thick,right] {$\infty_R$};
        \draw[blue!50!green, ultra thick,rounded corners] (-4,2) to[out=40,in=190] (-2,3.2) to[out=-10,in=200] node[midway,above,thick] {$\text{I}$} (1.2,2.4) node[above,black!100] {$P$};   
        \draw[orange!70!red, ultra thick,rounded corners] (2.3,2.5) node[above,black!100] {$T$} to[out=-3,in=145] node[midway, above, thick] {$\text{Rad}$} (4,2); 
        \filldraw[thick] (2.3,2.5) circle (0.05);
        \filldraw[thick] (1.2,2.4) circle (0.05);
        \draw[thick, dashed, opacity=1] (0,-2)  -- (2,0) --  (4,2) -- (2,4)   -- (0,6) -- (-2,4)  --  (-4,2)  -- ( -2,0)   --  (0,-2);
    \end{tikzpicture}
    \subcaption{}
    \label{fig:IslRad_Lorentz}
  \end{minipage}
  \begin{minipage}[b]{0.45\linewidth}
    \centering
    \begin{tikzpicture}[scale=1.2]
        \draw[thick, black!100] (-2,-1.8) -- (3,-1.8) -- (3,1.8) -- (-2,1.8) -- (-2,-1.8);
        \draw[thick, teal!80](-0.2,0) circle [radius=1.01];
        \filldraw[cyan!30, opacity=0.4,thick] (-0.2,0) circle [radius=1];
        
        \draw[orange!70!red, very thick, decorate, decoration={snake,amplitude=.4mm,segment length=2mm,post length=1mm}]
        (2,0) node [black, below, font=\normalsize]{$T$} -- (3,0);
        \draw[blue!50!green, very thick, decorate,decoration={snake,amplitude=.4mm,segment length=2mm,post length=1mm}] (-2,0) -- (0,0) node [black,below,font = \normalsize]{$P$};
        \draw (2.5,1.2) node[left,font = \normalsize]{Flat};
        \draw (-0.2,0.5) node[font=\normalsize]{AdS$_2$};
        \draw (2,0.3) node[orange!70!red,font=\normalsize,below,right]{\text{Rad}};
        \draw (-1.6,0.3) node[blue!50!green,font=\normalsize,below,right]{\text{I}};
        \filldraw[thick] (0,0) circle (0.05);
        \filldraw[thick] (2,0) circle (0.05);
    \end{tikzpicture}
    \subcaption{}
    \label{fig:IslRad_Euclid}
  \end{minipage}
  \caption{(a) The Lorentzian setup of the island configuration with the region $\text{Rad}\cup \text{I} =(\infty_L, P] \cup [T, \infty_R)$.
  The island region \text{I} purifies the entanglement between the left and right Rindler wedges and makes the entanglement entropy of \text{Rad} finite. (b) The Euclidean configuration where the Hartle-Hawking vacuum is prepared for the AdS$_2$ black hole plus flat bath. Since it is a pure state in the whole space, the entanglement entropy of $\text{Rad}\cup \text{I}$ equals that of the complementary region $\overline{\text{Rad}\cup \text{I}} = [P,T]$.}
  \label{fig:IslRad}
\end{figure}

Now we probe the Hawking radiation from the AdS$_2$ black hole by the entanglement entropy of a semi-infinite line $\text{Rad}=[T, \infty_R)$ in the right flat bath as in figure \ref{fig:Rad}. 
In what follows we consider a single-end island configuration $\mathrm{I}=(\infty_L, P]$ extending from the right Rindler wedge of the AdS$_2$ to the left flat bath as in the left panel of figure \ref{fig:IslRad}. 

We employ the island formula \eqref{island_formula} to find the dominant saddle in the gravitational path integral.
In this setup, the entanglement entropy of the region \text{Rad} suffers from an infrared divergence due to the infinite length. On the other hand, when we consider the entanglement entropy of the region $\text{Rad}\cup\text{I}$, the island region purifies the Hawking quanta in the right Rindler wedge, and the entropy is not affected by the infrared divergence.
Thus, the replica wormhole solution in the gravity region has less entanglement entropy than the black hole solution without an island region, always being the dominant saddle in the path integral.
We confirm the dominance of the island solution at the high temperature explicitly in this section.

For the island configuration the position $P$ is fixed by solving the QES condition \eqref{eq:QESn1}.
To this end it is convenient to parametrize the endpoints $P$ and $T$ by
\begin{align}
    P: y_R^\pm = t_a \mp a\ ,\qquad
    T: y_R^\pm = t_b \pm b\ ,
\end{align}
as shown in figure \ref{fig:IslRad}.
In the $W^\pm$ coordinates, they are written by
\begin{align}
    P: W_P^\pm = \pm e^{\pm \frac{2\pi}{\beta}(t_a \mp a)}\ ,\qquad
    T: W_T^\pm = \pm e^{\pm \frac{2\pi}{\beta}(t_b \pm b)}\ .
\end{align}

Once the island region is formed $S_\text{mat}$ is given by the entanglement entropy for the region $\text{Rad}\cup \text{I} =(\infty_L, P] \cup [T, \infty_R)$, but it equals the entropy of the complementary region $\overline{\text{Rad}\cup \text{I}} = [P,T]$ as the vacuum state on the Cauchy surface traversing the left and right AdS and bath regions is pure.
Since the spacetime is conformally flat the entanglement entropy of the matter follows from the flat space result \cite{Calabrese:2004eu}:\footnote{As explained below \eqref{eq:Renyi welding}, the conformal welding problem does not affect the entanglement entropy as the welding map becomes trivial when $n=1$.}
\begin{align}\label{ent_mat_oneint}
    S_\text{mat} = \frac{c}{6}\,\log \left(\frac{l^2}{\varepsilon^2\; \Omega_P \; \Omega_T}\right)\ ,
\end{align}
where we introduced the UV cutoff $\varepsilon$, which is independent of the spacetime cutoff $\epsilon$, and the length of the interval $l$ is given by
\begin{align}
    l^2 =\left| e^{\frac{4\pi b}{\beta}}+e^{-\frac{4\pi a}{\beta}}-e^{\frac{2\pi}{\beta}(t_b+b-t_a-a)}-e^{\frac{2\pi}{\beta}(t_a-a-t_b+b)}\right|\ .
\end{align}
The conformal factors at the endpoint $P$ and $T$ can be read off from \eqref{ConformalFactors} as
\begin{align}
    \Omega_P = \frac{1- e^{-\frac{4\pi}{\beta}a}}{2}\ ,\qquad
    \Omega_T = \frac{2\pi\epsilon}{\beta}\,e^{\frac{2\pi b}{\beta}}\ .
\end{align}
To fix the location of the island we extremize the generalized entropy in \eqref{island_formula} with respect to $P$.
Since the conformal factor and the area term do not depend on $t_a$, the QES condition \eqref{eq:QES} with respect to $t_a$ becomes
\begin{align}
    \begin{aligned}
    \sinh \frac{2\pi(t_a-t_b)}{\beta} = 0\ ,
    \end{aligned}
\end{align}
which yields $t_a = t_b$.
Then the generalized entropy reduces to
    \begin{align}
        \begin{aligned}
        \label{eq:gen1int}
            S_\text{island} = S_0 + \frac{2\pi}{\beta}\frac{\phi_r}{\tanh \frac{2\pi a}{\beta}} + \frac{c}{6}\log\left[\frac{2\beta}{\pi\varepsilon^2\epsilon}\frac{\sinh^2\frac{\pi(a+b)}{\beta}}{\sinh\frac{2\pi a}{\beta}}\right] \ .
        \end{aligned}
    \end{align}
Further extremizing $S$ over $a$ determines the island configuration:
\begin{align}
        \label{eq:QESresult}
    \partial_a S_\text{island} = 0\quad
    \rightarrow \quad
            \frac{1}{\sinh \frac{2\pi a}{\beta}} = \frac{\beta c}{12 \pi \phi_r}\frac{\sinh \frac{\pi(a-b)}{\beta}}{\sinh \frac{\pi(a+b)}{\beta}}\ .
\end{align}
In the high temperature limit where the effective gravitational coupling is weak
\begin{align}
    \kappa = \frac{\beta c}{24\pi \phi_r} \ll 1 \ ,
\end{align}
the QES condition \eqref{eq:QESresult} can be solved by $a= \infty$ at the leading order for $\kappa \ll 1$.
This means that the endpoint $P$ of the island region is  dynamically chosen to be the center of the AdS$_2$ space in the high temperature limit, and we find the generalized entropy for the solution with the island:
\begin{align}\label{eq:EntropyHighT}
    \begin{aligned}
        S_\text{island} 
         &=
         S_0 + \frac{2\pi\phi_r}{\beta} + \frac{c}{6}\log\left[\frac{\beta}{\pi\varepsilon^2\epsilon}\,e^{\frac{2\pi b}{\beta}}\right]\\
         &\approx
         S_0 + \frac{2\pi\phi_r}{\beta} + \frac{\pi c}{3\beta}\,b
         \ ,
    \end{aligned}
\end{align}
where we drop off the UV cutoff dependence and make the approximation in the second equality, which is valid for $b\gtrsim\frac{\beta}{2\pi}$ in the high temperature limit.

The entanglement entropy for the solution without the island is given just by the matter entropy similarly to \eqref{ent_mat_oneint}:
\begin{align}
    S_\text{no-island} = \frac{c}{3}\,\log \left[ \frac{\beta}{2\pi\varepsilon\epsilon}\,e^\frac{\pi\Lambda}{\beta}\right] \ ,
\end{align}
where we introduced the IR cutoff $\infty_R = (t_b, \sigma_R =\Lambda\gg b)$ for the radiation region.
This is always larger than \eqref{eq:gen1int} due to the IR divergence, so the island solution has less entropy and is favored in the high temperature limit.
    
\subsection{Capacity of entanglement}
\label{ss:capacity_ads2}

In section \ref{ss:entropy} we found the endpoint ($P$) of the island is fixed at the center of the AdS$_2$ by the QES condition in the high temperature limit ($\kappa \to 0$).
It follows from \eqref{eq:generalcap} that the capacity of entanglement is given by
\begin{align}\label{capacity_ads2}
    C = - \partial_n \phi^{(n)}(P)|_{n=1} + C_\text{mat}\ .
\end{align}
To evaluate the two terms that appear on the right hand side of this expression it is necessary to fix both the welding maps for the matter part and the $n$ dependence of the dilaton solution on the geometry with the island region, whose Euclidean geometry is given by the replica wormhole with a branch cut extending from the center $P$ of the AdS$_2$ to the point $T$ in the flat region (see figure \ref{fig:euclideanset}).
The first problem can be solved in the high temperature limit where the welding map trivializes for any $n$ \cite{Almheiri:2019qdq}.\footnote{We only keep the leading term in the $\kappa\to 0$ limit while \cite{Almheiri:2019qdq} includes the subleading term.}
Furthermore, in the same limit, the second problem simplifies as the Euclidean replica geometry has a conical singularity at the center of a disk on which the equation of motion for the dilaton can be solved analytically by imposing the asymptotic boundary condition on the boundary.
We exploit the high temperature limit to trivialize the welding maps and concentrate on the second problem below.

We begin with setting up the Euclidean geometry of the $\mathrm{AdS}{}_2$ region with $n=1$, which is related to the Lorentzian counterpart \eqref{eq:Lorentzian AdS2 BH} through the Wick rotation $t_{L}\mapsto -\i\, \theta$:
\begin{align}\label{EAdS2_metric}
    \d s^2_{\mathrm{ads}}\big|_{n=1} = \left(\frac{2\pi}{\beta}\right)^2 \frac{\d\theta^2 +\d\sigma^2}{\sinh^2 \left(\frac{2\pi\sigma}{\beta}\right)}\ , \qquad 
    \phi^{(n=1)} = -\frac{2\pi}{\beta} \frac{\phi_r}{\tanh \left(\frac{2\pi\sigma}{\beta}\right)}\ , 
\end{align}
where $\theta \sim \theta + \beta$, $-\infty < \sigma < -\epsilon$ and $\epsilon$ is a regularization of the asymptotic boundary.
Note that the left and right Rindler wedges are glued by the Euclidean time evolution by $\beta/2$.
To envision the replica calculation we introduce the complex coordinates $w, \bar w$ by $w=e^{\frac{2\pi}{\beta}(\sigma+\i\theta)}$.
Then the metric and the dilaton take the forms:
\begin{align}
    \d s^2_{\mathrm{ads}}\big|_{n=1}=\frac{4|\d w|}{(1-|w|^2)^2}\ ,\qquad \phi^{(n=1)}=\frac{2\pi\phi_r}{\beta}\frac{1+|w|^2}{1-|w|^2}\ ,
\end{align}
with $|w|\leq e^{-\frac{2\pi\epsilon}{\beta}}$.
Hence, the topology of the Euclidean AdS$_2$ black hole is a hyperbolic disk of circumference $2\pi$ and the dilaton field lives on the background. 

Now we focus on the high temperature limit where the island has the endpoint at the center of the AdS$_2$ (see section \ref{ss:entropy}).
In Euclidean signature, the endpoint is at the origin $w=0$ in the $w$ coordinates, so it is enough to solve the dilaton equation of motion on the hyperbolic disk with a conical singularity at the origin in order to evaluate the first term in the right hand side of \eqref{capacity_ads2} .

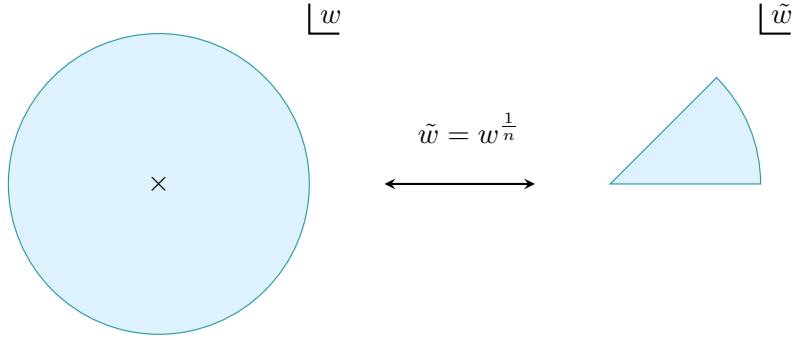
\begin{figure}[t]
    \centering
    \begin{tikzpicture}[scale=2]
    \filldraw[cyan!30, opacity=0.4,thick] (0,0) circle [radius=1cm]; 
    \draw[teal!80] (0,0) circle [radius=1cm];

    \draw[thick,black] (0,0) node [thick,black]{×};
    \draw[thick,black] (1.2,1.0)--(1.0,1.0) node[thick,black, above right] {$w$}--(1.0,1.2);
    \filldraw [cyan!30, opacity=0.4,thick] (4,0)  arc (0:45:1) --(3,0)--cycle;
        \draw [teal!80] (4,0)  arc (0:45:1) --(3,0)--cycle;

    \draw[thick,black] (4.2,1.0)--(4.0,1.0) node[thick,black,above right] {$\tilde{w}$}--(4.0,1.2);

    \draw[thick,<->,>=stealth](1.5,0)--(2.5,0);
    \draw (1.65,0.2) node[above right,thick,black,text width=2cm]{$\tilde{w} = w^{\frac{1}{n}}$};
\end{tikzpicture}
    \caption{Two ways of describing a conical singularity. In the $w=e^{\frac{2\pi}{\beta}(\sigma+\i\theta)}$ coordinates, the orbifold $\CM_n$ has an $n$ dependent metric with the inverse temperature $\beta$. On the other coordinates $\tilde{w} = e^{\frac{2\pi}{\beta}(\tilde{\sigma}+\i\tilde{\theta})}$, the geometry is uniformized and its metric realizes an AdS$_2$ disk with the inverse temperature $\beta/n$.}
    \label{fig:conical}
\end{figure}

We proceed with the replica geometry on the orbifold $\CM_n$ where the conical singularity is introduced so that the metric has an explicit $n$ dependence while keeping the periodicity of the Euclidean time $\theta \sim \theta + \beta$.
To determine the $n$ dependent dilaton profile in the presence of the conical singularity, it is convenient to use another equivalent description on the geometry (see figure \ref{fig:conical}), which is obtained by following uniformization of the coordinates
\begin{align}
     \label{eq:unif}
     \tilde{w} = w^{\frac{1}{n}}\ .
\end{align}
After the uniformization, the metric in the $\tilde{w}$ coordinates becomes independent of $n$:
 \begin{align}\label{tilde_metric}
     \d s^2_{\mathrm{ads}} = \frac{4|\d\tilde{w}|^2}{(1-|\tilde{w}|^2)^2}\ ,
 \end{align}
but the Euclidean time for the tilde coordinates defined through $\tilde{w}=e^{\frac{2\pi}{\beta}(\tilde{\sigma}+\i\tilde{\theta})}$ has the $n$ dependent periodicity $\tilde{\theta}\sim \tilde{\theta}+\beta/n$.

The dilaton equation of motion can be solved straightforwardly in the $\tilde{w}$ coordinates as
\begin{align}
    \phi = \frac{2\pi\tilde{\phi}_r}{\beta}\, \frac{1+|\tilde{w}|^2}{1-|\tilde{w}|^2} 
\end{align}
by imposing the boundary condition 
\begin{align}
    \begin{aligned}
    \phi|_{\rm bdy} = \frac{\tilde{\phi}_r}{\tilde{\epsilon}}\ ,\qquad
    \tilde{w} = e^{\frac{2\pi}{\beta}(-\tilde{\epsilon}+\i\tilde{\theta})}\ ,\qquad
    \tilde{\epsilon} \to 0 \ ,
    \end{aligned}
\end{align}
where $\tilde{\phi}_r$ is the renormalized dilaton in the $\tilde{w}$ coordinates.
To find the dilaton solution in the $w$ coordinates we impose a similar boundary condition
\begin{align}\label{dilaton_asympt_bc}
    \phi|_{\rm bdy} = \frac{\phi_r}{\epsilon}\ ,\qquad
    w = e^{\frac{2\pi}{\beta}(-\epsilon+\i\theta)}\ ,\qquad
    \epsilon \to 0 \ .
\end{align}
The conformal map \eqref{eq:unif} between the  $w$ and $\tilde{w}$ coordinates reduces to the relation between the cutoffs $\tilde{\epsilon} = \epsilon/n$.
Then the comparison between the two asymptotic boundary conditions leads to the relation $\tilde{\phi}_r = \phi_r/n$.
Since the dilaton is a scalar field, its transformation law is trivial.
As a result, we obtain the dilaton solution in the presence of conical singularity at the origin:
\begin{align}
    \label{eq:finitedilaton}
    \phi^{(n)} = \frac{2\pi\phi_r}{n\beta}\,\frac{1+|w|^{\frac{2}{n}}}{1-|w|^{\frac{2}{n}}} =-\frac{2\pi}{n\beta}\: \frac{\phi_r}{\tanh \frac{2\pi\sigma}{n \beta}}\ .
\end{align}
In appendix \ref{sec:local}, we verify that this expression reproduces the local solution around the conical singularity and near the disk boundary at the leading order of $n-1$ by solving the dilaton equation on the $w$ coordinates.

Now we are ready to compute the capacity of entanglement by the formula \eqref{eq:generalcap}.
It follows from \eqref{eq:finitedilaton}
\begin{align}
        -\partial_n \phi^{(n)}\big|_{n=1} = \frac{2\pi\phi_r}{\beta} \qquad (a\to \infty) \ ,
\end{align}
where we fix the location of the conical singularity by the QES condition as in section \ref{ss:entropy}.
For conformal matters, the replica partition function with a single interval takes the form:
    \begin{align}
        \label{eq:spectrumRenyi}
         \log Z_\text{CFT}[\widetilde \CM_n]
         = \left(\frac{l^2}{\varepsilon^2\;\Omega_P\; \Omega_T}\right)^{-\Delta_n},\qquad
        \Delta_n = \frac{c}{12}\left(n-\frac{1}{n}\right) \ ,
    \end{align}
where we drop off the $n$ dependence of the welding maps, which is valid in the $\kappa\to 0$ limit.
Thus the capacity for matter CFT equals to the entanglement entropy \eqref{ent_mat_oneint}.
In total the capacity for the island phase in the high temperature limit becomes
\begin{align}\label{eq:CapacityHighT}
        C  \approx  \frac{2\pi\phi_r}{\beta}+\frac{\pi c}{3\beta}\,b \ ,
\end{align}
where we drop the UV cutoff dependence.
For a consistency check we rederive the same result from the gravitational path integral on the boundary curve in appendix \ref{ss:fromreplica}.

\subsection{Relation to thermodynamic quantities}\label{ss:thermal}

We have demonstrated the calculations of several quantum information measures for the replica wormhole in the familiar model of Hawking radiation.
To gain some physical intuition behind the results and examine the validity of our formula \eqref{eq:generalcap} for the capacity of entanglement, 
we will discuss their relations to thermodynamic quantities, the coarse-grained or black holes entropy and capacity in the current model. 

\begin{figure}
    \centering
    \begin{tikzpicture}[scale=0.6]
        \draw[fill=cyan!30, opacity=0.4] (0,0) circle [x radius=1cm, y radius=2cm]; 
        \draw[teal!80] (0,0) circle [x radius=1cm, y radius=2cm]; 
        \draw[dashed,thin,opacity=0.5] (4,0) circle [x radius=1cm, y radius=2cm];
        \draw (0,2)--(6,2) (0,-2)--(6,-2);
        \draw (4.5,1.5) circle [x radius=0.4cm, y radius=0.2cm];
        \fill[fill=gray!30,opacity=0.4] (4.5,1.5) circle [x radius=0.4cm, y radius=0.15cm];
        \draw (0,2) node[above] {\small$\sigma=0$} (4,2) node[above] {\small $\sigma=b$};
        \draw(0,0) node[rotate=-9]{\scriptsize AdS$_2$} (5.8,-1.8) node[above]{\scriptsize Flat};
        \draw[snake it] (0.6,1.5)--(4.1,1.5);
        \filldraw (4.5,1.5) circle [radius=1pt];
        \draw (4.8,1.4) node[right] {\small $T$};
        \draw[-latex] (1.5,1) arc [start angle=30, end angle=-30, x radius=1cm, y radius=2cm];
        \draw[-latex] (1.2,0.8)--(2.5,0.8) node[below]{\small $\sigma$};
        \draw (1.63,-0.85) node[right,fill=white]{\small $\tau \sim \tau + \beta$};
        \draw[-latex](7-0.5,0)--(7+2.5,0);
        \draw(8,0) node[align=center]{\scriptsize Conformal \\ \scriptsize mapping};
    \end{tikzpicture}
    \begin{tikzpicture}[scale=0.6]
        \draw[fill=cyan!30, opacity=0.4] (0,0) circle [x radius=1cm, y radius=2cm]; 
        \draw[teal!80] (0,0) circle [x radius=1cm, y radius=2cm]; 
        \fill[fill=gray!30,opacity=0.4] (4,0) circle [x radius=1cm, y radius=2cm];
        \draw[dashed,thin,opacity=0.5] (4,2) arc [start angle=90, end angle=270, x radius=1cm, y radius=2cm];
        \draw (4,-2) arc [start angle=270, end angle=450, x radius=1cm, y radius=2cm];
        \draw[decorate,decoration={snake,amplitude=.4mm,segment length=2mm,post length=1mm}] (0,2)--(4,2);
        \draw (0,-2)--(4,-2);
        \draw[latex-latex] (0,2.2)--(4,2.2);
        \draw(0,0) node[rotate=-9]{\scriptsize AdS$_2$} (2.5,-1.8) node[above]{\scriptsize Flat};
        \draw(2,2.1) node[above]{$b$};
        \filldraw (4,0) node[below] {\small $T$} circle [radius=1pt];
    \end{tikzpicture}
    \caption{The Euclidean geometry of the model is the semi-infinite cylinder anchored on the boundary of the gravitating disk region [Left]. It can be conformally mapped to the finite cylinder geometry [Right], which is used to calculate the thermal free energy.}
    \label{fig:cylinder}
\end{figure}
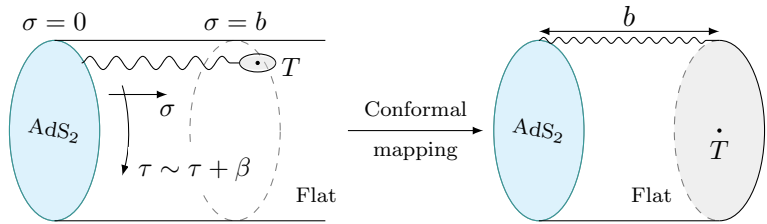

The Euclidean geometry of the model in section \ref{ss:LorentzianAdS2} consists of a disk region with dynamical gravity and a cylinder region without gravity  (see the left panel of figure \ref{fig:cylinder}).
The former is the Euclidean AdS$_2$ black hole of inverse temperature $\beta$ with the metric \eqref{EAdS2_metric} and the latter is the flat bath region. 
On this background, we can estimate the thermal free energy semiclassically 
by using the on-shell value of the JT gravity action \eqref{eq:action} and the thermal partition function of the matter:
\begin{align}
        I(\beta)
        &\simeq I_{\text{grav}}^{\text{on-shell}}(\beta) - \log Z_{\text{CFT}}(\beta)\ . 
\end{align}
To interpret the results \eqref{eq:EntropyHighT} and \eqref{eq:CapacityHighT} as thermodynamic quantities, we will consider the disk and cylinder regions separately. 
On the disk region, the on-shell gravity action is evaluated with a proper counterterm as
\begin{align}
    \begin{aligned}
     I_{\rm grav}^{\text{on-shell}}(\beta)
        &= - S_0 - \frac{\phi_b}{2\pi}\int_{0}^{\beta}\d\tau \sqrt{\gamma}\, (\CK -1) \\  
        &= -S_0 - \frac{\pi\phi_r}{\beta} + O(\epsilon)\ ,
     \end{aligned}
\end{align}
where  $\phi_b$ is the boundary value of the dilaton, $\phi_b\equiv \phi_r/\epsilon$ and $\gamma$ is the induced metric on the boundary circle at $\sigma = -\epsilon$.
For conformal matter on the disk, we see from dimensional analysis that the partition function should be a function of the dimensionless constant $\beta/\epsilon$ and does not depend on the size of the region $b$.
It is either zero or UV divergent, so may be canceled by adding an appropriate counterterm. 
We do not take this contribution into account in the following consideration.

Next let us move to the flat bath region without gravity, which is the semi-infinite cylinder whose boundary is glued to the boundary circle of the disk (the left panel of figure \ref{fig:cylinder}). 
In the replica calculation of the entanglement entropy, we probe the Hawking radiation through the radiation region extending from a point $T$ to the infinity.
In the Euclidean picture, the semi-infinite cylinder has the same radiation region extending from the point $T$ where a conical singularity arises in the replica calculation.
Having the comparison with the results in the previous subsection in mind, we consider the semi-infinite cylinder with a small hole around the point $T$.
This configuration is conformally equivalent to a finite cylinder (the right panel of figure \ref{fig:cylinder}), so we calculate the matter partition function on the latter with circumference $\beta$ and length $b$.
The matter partition function on the cylinder becomes
\begin{align}
    \log Z_{\text{CFT}}(\beta) 
    \simeq \frac{\pi c}{6\beta}\, b\ ,
\end{align} 
in the high temperature limit ($\beta \to 0$) up to boundary entropy contributions that are independent of $\beta$ \cite{Affleck:1991tk}. 
Combining the disk and cylinder contributions, we find the thermal entropy $S_{\text{th}}(\beta)$ and capacity $C_{\text{th}}(\beta)$ in the high temperature limit:\footnote{For the derivation from the boundary curve viewpoint in the JT gravity, see \cite{Maldacena:2016upp}.}
\begin{align} 
    S_{\text{th}}(\beta) \label{eq:thermalS}
    &\equiv  (\beta\partial_{\beta} -1) I(\beta)
    \simeq S_0 + \frac{2\pi \phi_r}{\beta} + \frac{\pi c}{3\beta}\,b\ , \\
    C_{\text{th}}(\beta) \label{eq:thermalC}
    &\equiv - \beta^2 \partial_{\beta}^2 I(\beta) 
    \simeq \frac{2\pi \phi_r}{\beta} + \frac{\pi c}{3\beta}\,b \ .
\end{align}
These precisely coincide with \eqref{eq:EntropyHighT} and \eqref{eq:CapacityHighT} derived from the island \eqref{island_formula} and capacity formula \eqref{eq:generalcap} respectively.

The reason for the coincidence between the quantum informational and thermodynamic quantities for the matter part may be understood as follows.
At the leading order in the high temperature limit, the entanglement entropy for the single interval on the semi-infinite cylinder matches the thermal entropy \eqref{eq:thermalS} as
\begin{align}
    S = \frac{c}{3}\log \left(\frac{\beta}{\pi}\sinh \frac{\pi b}{\beta}\right) \approx \frac{\pi c}{3\beta}\, b \ .
\end{align}
A similar consideration also works for the capacity, explaining why the two different quantities match in the high temperature limit.

Let us expand on the relations for the gravity part in the island phase at high temperature.
In the island or replica wormhole phase, all of the $n$ replicated disk boundaries are fused to one disk with circumference $n \beta$.
Thus it is expected that the replica wormhole is described by the black hole of inverse temperature $n\beta$.
Under this assumption the refined R\'{e}nyi entropies in the high temperature equals the thermal entropy with inverse temperature $n \beta$:
\begin{align}
    \tilde{S}^{(n)}_{\text{island}}
    \approx \partial_{1/n} \left( \frac{1}{n} \log \frac{Z(n \beta)}{(Z(\beta))^n}\right)
    = S_{\text{th}} (n \beta) \ .
\end{align}
It follows that the entropy and capacity in the island phase coincide with the thermal quantities \eqref{eq:thermalS} and \eqref{eq:thermalC}:
\begin{align}
    S_{\text{island}} = S_{\text{th}} (\beta)\ , \qquad
    C_{\text{island}} = C_{\text{th}} (\beta)\ . 
\end{align}
These coarse-grained values reproduce the results \eqref{eq:EntropyHighT} and \eqref{eq:CapacityHighT} derived from the formulas \eqref{island_formula} and \eqref{eq:generalcap}. 
The coincidence implies that the classical or thermal mixture for the  $n$ replicas of CFTs via gravity plays a key role in the island phase.

Here we observe the coincidence between the entanglement entropy and the thermal entropy, the capacity of entanglement and the heat capacity.
A similar observation to this subsection was also made in the end-of-the-world brane model \cite{Kawabata:2021hac}. Reproducing the thermodynamic counterpart can be seen as giving a piece of evidence for the validity of our formula \eqref{eq:generalcap}.

\section{Discussion}\label{ss:discussion}

In this paper, we discussed the capacity of entanglement in a two-dimensional dilaton gravity coupled to conformal matter with a large central charge, 
which is the setup that the island formula of the entanglement entropy for the Hawking radiation was proposed  \cite{Almheiri:2019hni,Penington:2019npb,Almheiri:2019psf}. 

In the setup, we proposed the formula \eqref{eq:generalcap} for the capacity of entanglement based on the replica wormhole prescription  \cite{Penington:2019kki,Almheiri:2019qdq}. 
Since the gravitational part of the capacity is essentially a derivative of the corresponding part of the entropy with respect to the replica parameter, 
the capacity has no topological term which the entanglement entropy has.
We argued based on the formula that the capacity shows a discontinuity at the Page time when the entropy smoothly transits to the island saddle from the black hole one. 

In contrast to the entanglement entropy, the capacity cannot be calculated without solving the conformal welding problem in general. 
To avoid the problem, we focused on a simple setting, the high temperature limit of an eternal AdS$_2$ black hole background in the JT gravity glued with an auxiliary Minkowski space as a heat bath.
We calculated the entropy \eqref{eq:EntropyHighT} and capacity \eqref{eq:CapacityHighT} in the island phase that always dominates in this limit and observed that they correspond to the thermodynamic quantities, which we expect to be a universal phenomenon in the late time of the Hawking radiation.

From the observer in the radiation system, what distinguishes the island and no-island phases is the difference of the classical mixture among the $n$-copies of CFTs triggered by the island or replica wormhole formation. 
The $n$-copies of CFTs are separated gravitationally and have no mixture in the no-island phase 
but they are thermally mixed in the island phase. 
If we add matter contributions in more general setups, this difference will cause a phase transition at the Page time when the Page curve of the entropy bends. 
We speculate that the classical mixture should be a source of the discontinuous jump \eqref{eq:capdiscont} of the capacity in the phase transition.

There are a number of issues that we hope to address in future.
The first investigation that comes to mind is to study the capacity in various backgrounds in which the island formula for the entropy was already discussed. 
While the conformal welding problem remains to be an obstacle to the analytic studies it would deserve to examine if the capacity becomes discontinuous in a tractable setup and verify our conjecture.

A more practical approach to the conformal welding problem is to implement the numerical calculation as is carried out in \cite{Mirbabayi:2020fyk} for the replica wormhole with $1< n \leq2$. 
The numerical analysis of the capacity will make an important cross-check of our results and expand them into general parameter regions in which the phase transition can happen. 

It is also interesting to consider the capacity in the doubly holographic setups. 
As discussed in \cite{Nakaguchi:2016zqi,deBoer:2018mzv}, 
the holographic dual of the capacity is described by the graviton fluctuation around the minimal codimension-two surface associated with the holographic entanglement entropy, but it should be generalized to the AdS/BCFT setup \cite{Takayanagi:2011zk,Fujita:2011fp} to incorporate the gravity region on the boundary theory.
In particular, it is worthwhile to understand how the welding problem that matters for the capacity is dualized in the holographic setup.

In this paper, we only discussed the leading gravitational contributions to the capacity from the semiclassical saddles. At the leading level, we expect that the capacity can probe the phase transition between the two saddles bridged via non-perturbative instanton corrections as the discontinuity at the Page time.
Beyond the semiclassical regime, however, there are $O\left(G_N^{-1/2}\right)$ corrections from graviton quantum fluctuation around the saddles. 
As shown in our previous work in the end-of-the-world-brane model \cite{Kawabata:2021hac}, the quantum correction can smooth out the discontinuity of the capacity and the phase transition reduces to a crossover.
The quantum corrected behavior of the capacity around the Page time may be accessible by applying the fixed area state prescription \cite{Akers:2018fow,Dong:2018seb,Marolf:2020vsi} to the current setup.

\acknowledgments
We are grateful to K.\,Goto, V.\,Hubeny, M.\,Rangamani and T.\,Ugajin for valuable discussions.
The work of T.\,N. was supported in part by the JSPS Grant-in-Aid for Scientific Research (C) No.19K03863, the JSPS Grant-in-Aid for Scientific Research (A) No.16H02182 and No.21H04469, and the JSPS Grant-in-Aid for Transformative Research Areas (A) No.\,21H05190. 
The work of K.\,W. was supported
by U.S. Department of Energy grant DE-SC0019480 under the HEP-QIS QuantISED program and by funds from the University of California. The works of K.\,K. and Y.\,O. were supported by Forefront Physics and Mathematics Program to Drive Transformation (FoPM), a World-leading Innovative Graduate Study (WINGS) Program, the University of Tokyo.
The work of Y.\,O. was also supported by JSPS fellowship for Young students and by JSR fellowship, the University of Tokyo.
\appendix

\section{Capacity of entanglement from replica wormhole}
\label{ss:fromreplica}
We used the island formula to determine the position of the conical singularity at the origin of AdS$_2$ spacetime in section \ref{ss:entropy}.
The aim of this appendix is to determine the location of the singularity and reproduce the capacity of entanglement from the boundary curve.
We focus on the single-end island in AdS$_2$ as in the previous analysis.

\subsection{Gravitational path integral on boundary curve}
\label{sss:schwarzian}
Below we work in Euclidean geometry, setting the inverse temperature of the AdS${}_2$ black hole to $\beta=2\pi$ for simplicity.\footnote{We will recover $\beta$ dependence from the dimensional analysis later.} We use the $w=e^{-r+\i\theta}$ coordinate inside the disk introduced in the section \ref{ss:capacity_ads2}.
The boundary of the disk is characterized by the coordinate $\theta$.
We parametrize the boundary curve by the time $\tau$ along the circle. 

We describe the JT gravity on the finite $n$ replica wormhole by the boundary mode $\theta(\tau)$.
Assuming the replica symmetry, the replica wormhole in the $n$-fold cover has $\BZ_n$-fixed points. After orbifolding by $\BZ_n$, we need the cosmic brane term that supplies the conical singularity at a generic point $w=A=e^{-a}$. In our setup, the relation \eqref{eq:frmcon} between the action of the replica wormhole $\widetilde{\CM}_n$ and its $\BZ_n$-orbifold ${\CM}_n$ becomes \cite{Almheiri:2019qdq}
\begin{align}
    -\frac{1}{n}I_{\rm grav}[\widetilde{\CM}_n] = \frac{S_0}{4\pi}\left[\int \CR + \int 2\CK \right] + \int\frac{\phi}{4\pi}(\CR + 2) + \frac{\phi_b}{4\pi}\int 2\CK - \left(1-\frac{1}{n}\right)[S_0+\phi(A)] \ ,
\end{align}
where the right hand side integrates over the orbifold ${\CM}_n$ except for the terms where the extrinsic curvature appears, which are integrated over the boundary $\partial \CM_n$.
Substituting the on-shell curvature into the action, we get the action for the boundary mode $\theta(\tau)$ \cite{Almheiri:2019qdq}:
\begin{equation}
\label{eq:onshell}
    -I_{\rm grav}(n) \equiv -I_{\rm grav}[\widetilde{\CM}_n]= S_0 + n\,\frac{\phi_r}{2\pi}\,\int^{2\pi}_0 \d\tau \left[\{e^{\i\theta},\tau\}+\frac{1}{2}\left(1-\frac{1}{n^2}\right)R(\theta)\right]\ ,
\end{equation}
with
\begin{align}
    R(\theta) = -\frac{(1-A^2)^2\,(\partial_\tau \theta)^2}{|1-Ae^{\i\theta}|^4}\ ,
\end{align}
where we used the relation between the boundary value of the dilaton $\phi_b$ and the renormalized dilaton $\phi_r$ in approaching the boundary $w=e^{-\epsilon+\i\theta}$: $\phi_b = \phi_r/\epsilon$.

We can treat the dynamics of the JT gravity by the boundary mode action \eqref{eq:onshell}. 
Now we want to consider the equation of motion for the boundary mode and vary the gravitational part as follows:
\begin{align}\label{eq:vari}
        -\frac{1}{n}\,\delta I_{\rm grav} = \frac{\phi_r}{2\pi}\int\d\tau \left[\delta\{e^{\i\theta},\tau \}+\frac{1}{2}\left(1-\frac{1}{n^2}\right)\delta R(\theta)\right]\ .
\end{align}
To derive the equation of motion for the time evolution with respect to $\tau$ we consider the variation  $\delta\theta(\tau)$ of the boundary mode $\theta$.
Then, the chain rule tells us that \eqref{eq:vari} becomes \cite{Almheiri:2019qdq}
\begin{align}\label{eq:fortau_vari}
        -\frac{1}{n}\,\delta I_{\rm grav} = \frac{\phi_r}{2\pi}\int\d\tau \left[\partial_\tau\{e^{\i\theta},\tau\}+\frac{1}{2}\left(1-\frac{1}{n^2}\right)\partial_\tau R(\theta)\right]\frac{\delta \theta}{\partial_\tau \theta}\ .
\end{align}
Note that in the parenthesis the ADM energy $M$ of the system appears
\begin{align}
    M = 
    -\frac{\phi_r}{2\pi}\left[\{e^{\i\theta},\tau\}+\frac{1}{2}\left(1-\frac{1}{n^2}\right)R(\theta)\right]\ .
\end{align}
The equation of motion including conformal matters leads to the conservation of energy between the ADM energy and the flux of matter fields \cite{Maldacena:2016upp}:
\begin{align}
    \frac{\phi_r}{2\pi}\left[\partial_\tau\{e^{\i\theta},\tau\}+\frac{1}{2}\left(1-\frac{1}{n^2}\right)\partial_\tau R(\theta)\right]
    =
    \i\,\left[T_{yy}(\i\tau)-T_{\bar{y}\bar{y}}(-\i\tau)\right]\ ,
\end{align}
where $T_{yy}$ is the stress tensor in the coordinate $y$ outside the disk.
The state preparation on the AdS$_2$ plus flat space requires the coordinates which holomorphically cover the whole space.
Then, we are forced to solve the conformal welding problem to get the stress tensor in the $y$ coordinates.
Assuming that the welding map outside can be written as $F(e^y)$, we get the boundary equation \cite{Almheiri:2019qdq}:
\begin{align}
    \partial_\tau\left[\{e^{\i\theta(\tau)},\tau\}+\frac{1}{2}\left(1-\frac{1}{n^2}\right)R(\theta(\tau))\right]
        =
        \i\, \kappa\, e^{2\i\tau}\left[-\frac{1}{2}\left(1-\frac{1}{n^2}\right)\frac{F'(e^{\i\tau})^2}{F(e^{\i\tau})^2}-\{F,e^{\i\tau}\}\right] + \text{c.c.}\ ,
\end{align}
where $\kappa = c\beta/(24\pi \phi_r) \propto \beta$ with recovering the $\beta$ dependence. 
In the high temperature limit $\kappa \to0$, we get the equation of motion
\begin{align}
    \begin{aligned}
    \label{eq:eombdy2}
    \partial_\tau\left[\{e^{\i\theta(\tau)},\tau\}+\frac{1}{2}\left(1-\frac{1}{n^2}\right)R(\theta(\tau))\right] = 0\ ,
    \end{aligned}
\end{align}
which implies the conservation of ADM energy of the AdS$_2$ spacetime.
In the next section, we solve this boundary equation at the order $O(n-1)$.

We are going to define another action for a later convenience:
\begin{equation}
\label{eq:orbaction}
    \hat{I}_n = \frac{1}{n}\,I_{\rm grav}(n) - I_{\rm grav}(1) - \frac{1}{n}\,\log Z_{\rm CFT}[\widetilde{\CM}_n] + \log Z_{\rm CFT}[\widetilde{\CM}_1] \ .
\end{equation}
This relates with the entanglement entropy and the capacity of entanglement through the derivative for $n$.
If we act the derivative with respect to $n$ on this action $\hat{I}_n$, then we get the entanglement entropy $S$ and the capacity of entanglement $C$ through the refined R\'{e}nyi entropy and $n^{th}$ capacity of entanglement \cite{Nakaguchi:2016zqi,Dong:2017xht}:
\begin{align}
\begin{aligned}
    \tilde{S}^{(n)} &= n^2\partial_n \hat{I}_n\quad &\to \qquad S &= \partial_n \hat{I}_n |_{n = 1}, \\
    C^{(n)} &= -n\partial_n(n^2\partial_n \hat{I}_n )\quad &\to \qquad C &= -\partial_n^2 \hat{I}_n\big|_{n=1} -2 \partial_n \hat{I}_n |_{n=1}.
    \label{eq:cap}
    \end{aligned}
\end{align}
We apply this formula to derive the generalized entropy and the capacity of entropy in gravity. 

Substituting the on-shell action \eqref{eq:onshell} to \eqref{eq:orbaction}, we obtain
\begin{align}
    \hat{I}_n = \hat{I}_{\rm grav}(n) + \hat{I}_{\rm mat}(n) \ ,
\end{align}
where we divide the action into two parts, the gravitational and matter sectors:
\begin{align}
    \hat{I}_{\rm grav}(n) &= \left(1-\frac{1}{n}\right)S_0 + \frac{\phi_r}{2\pi}\int^{2\pi}_0 \d\tau \left[\{e^{\i\theta},\tau\}\big|_{n=1}-\{e^{\i\theta},\tau\}\big|_{n}\right]
    -\frac{\phi_r}{4\pi}\left(1-\frac{1}{n^2}\right)\int^{2\pi}_{0}\d \tau R(\theta)\ , \\
    \hat{I}_{\rm mat}(n) &= - \frac{1}{n}\,\log Z_{\rm CFT}[\widetilde{\CM}_n] + \log Z_{\rm CFT}[\widetilde{\CM}_1]\ .
\end{align}

We reproduce the generalized entropy following \cite{Almheiri:2019qdq}.
First, we act the derivative with respect to $n$ on the action $\hat{I}_n$:
\begin{align}
\label{eq:1grav}
    \partial_n \hat{I}_{\rm grav}(n) &= \frac{1}{n^2}\,S_0 - \frac{\phi_r}{2\pi}\int^{2\pi}_{0}\d\tau\, \partial_n\{e^{\i\theta},\tau\}|_n - \frac{\phi_r}{2\pi n^3}\int^{2\pi}_0\d\tau R(\theta) - \frac{\phi_r}{4\pi}\left(1-\frac{1}{n^2}\right)\int^{2\pi}_0\d\tau\, \partial_n R(\theta) \ , \\
    \partial_n \hat{I}_{\rm mat}(n) &= \frac{1}{n^2}\,\log Z_{\rm CFT}[\widetilde{\CM}_n]-\frac{1}{n}\,\partial_n \log Z_{\rm CFT}[\widetilde{\CM}_n]\big|_g \ ,
    \label{eq:1mat}
\end{align}
where the metric on the $n$-fold cover $\widetilde{\CM}_n$ is fixed to AdS$_2$, so \eqref{eq:1mat} does not contain any $n$ dependence through the metric of $\widetilde{\CM}_n$.
On the other hand, \eqref{eq:1mat} has the $n$ dependence of the conformal welding map because we evaluate the matter entropy on the geometry where the conformal welding is implemented in order to prepare the state for CFT.

\eqref{eq:1grav} and \eqref{eq:1mat} leads to the refined R\'{e}nyi entropy:
\begin{align}
\begin{aligned}
\label{eq:n1}
    \Tilde{S}^{(n)} 
    &= n^2\partial_n \hat{I}_n \\
    &= S_0 - \frac{\phi_r}{2\pi n}\int^{2\pi}_0\d\tau R(\theta) +  \log Z_{\rm CFT}[\widetilde{\CM}_n]
    - n\,\partial_n \log Z_{\rm CFT}[\widetilde{\CM}_n]\big|_{g}\\ 
    &\;\;\;\;\;\;\;- n^2\,\frac{\phi_r\, }{2\pi}\int^{2\pi}_{0}\d\tau\, \partial_n\{e^{\i\theta},\tau\} - \frac{\phi_r}{4\pi}\,(n^2-1)\int^{2\pi}_0\d\tau\, \partial_n R(\theta) \\
    &= S_0 - \frac{\phi_r}{2\pi n}\int^{2\pi}_0\d\tau\, R(\theta) +  \Tilde{S}^{(n)}_{\rm mat} \ .
\end{aligned}
\end{align}
The second line reduces to the third line by the equation of motion for the boundary mode in the high temperature limit, which is obtained by the same procedure as \eqref{eq:fortau_vari} but with varying $\theta$ with respect to $n$ while fixing $\tau$.

We proceed onto the capacity of entanglement.
All we have to do is to act one more derivative with respect to $n$ on terms \eqref{eq:n1}.
Then, the capacity of entanglement is
\begin{align}
    \begin{aligned}
    \label{eq:capboundary}
    C &= -n \frac{\partial \tilde{S}^{(n)}}{\partial n}\bigg|_{n=1} \\
    &= -\frac{\phi_r}{2\pi}\int^{2\pi}_0\d\tau\, R(\theta)|_{n=1}
    + \frac{\phi_r}{2\pi}\int^{2\pi}_0\d\tau \,\partial_n R(\theta)\big|_{n=1} + C_{\rm mat}
    \end{aligned}
\end{align}
where we use $C_{\rm mat}=-n\partial_n \tilde{S}^{(n)}_{\rm mat}\big|_{n=1}$. The second term implies that we have to solve the boundary mode $\theta(\tau)$ at the order $O(n-1)$.

We have derived the entanglement entropy and the capacity of entanglement in terms of the gravitational path integral in the on-shell Schwarzian action.
The capacity of entanglement is shown to depend on the boundary curve of the order $O(n-1)$.
Next, we solve the equation of motion for the boundary mode and obtain a consistent answer with the previous formula of the capacity of entanglement.

\subsection{Replica solution as $n\to1$}
\label{sss:repsolution}
We solve the equation of motion for the boundary mode
\begin{align}\label{eq:eombdy}
    \partial_\tau\left[\{e^{\i\theta},\tau\}+\frac{1}{2}\left(1-\frac{1}{n^2}\right)R(\theta)\right] = 0\ ,
\end{align}
where the right hand side is vanishing in the high temperature limit.
The capacity of entanglement \eqref{eq:capboundary} needs the solution of the order $O(n-1)$, so we perturbatively expand the boundary mode $\theta(\tau)$ near $n=1$:
\begin{align}
    \theta(\tau) = \tau + (n-1)\, \delta\theta(\tau) + O\left((n-1)^2\right) \ .
\end{align}
Then, the equation of motion \eqref{eq:eombdy} boils down to 
\begin{align}\label{eq:bdyeq}
    \partial_\tau^2(\partial_\tau^2 + 1)\, \delta\theta(\tau) = \frac{4A(1-A^2)^2\sin\tau}{|1-Ae^{\i\tau}|^6}\ .
\end{align}
Now we impose some conditions for the boundary mode $\theta(\tau)$:
\begin{itemize}
    \item periodicity: \qquad $\theta(\tau+2\pi) = \theta(\tau) + 2\pi$
    $\rightarrow$ $\delta\theta(\tau+2\pi) = \delta \theta(\tau)$\ .
    \item odd function: \qquad $\theta(\tau) = -\theta(-\tau)$ $\rightarrow$ $\delta\theta(\tau) = -\delta\theta(-\tau)$\ .
\end{itemize}
In order to solve the boundary equation \eqref{eq:bdyeq}, we expand $\delta\theta$ by the Fourier modes that satisfy the above conditions:
\begin{align}
    \delta \theta = \sum_{m>0} c_m \sin (m\tau)\ .
\end{align}
With this, the right hand side of the equation \eqref{eq:bdyeq} can be expanded as
\begin{align}
    \frac{4A(1-A^2)^2\sin\tau}{|1-Ae^{\i\tau}|^6} = \sum_{m>0} \frac{2\, m A^m(1+A^2(1-m)+m)}{1-A^2}\, \sin (m\tau)\ .
\end{align}
Then, the equation of motion is
\begin{align}
    \sum_{m>0}c_m m^2(1-m^2)\sin(m\tau) = \sum_{m>0} \frac{2\,m A^m(1+A^2(1-m)+m)}{1 - A^2}\, \sin (m\tau)\ .
\end{align}
The important point is that the left hand side for $m=1$ is vanishing regardless of the coefficient $c_m$, but the right hand side for $m=1$ has the Fourier coefficient $2A/(1-A^2)$.
In order to satisfy the boundary equation, we have to impose
\begin{align}
    \frac{2A}{1-A^2}\sin(\tau)=0\ .
\end{align}
Then, the equation of motion for the boundary mode yields that the position of the conical singularity must be $A=0$, the origin of AdS$_2$. This is consistent with the previous result derived from the QES condition.

To get the formula for the capacity of entanglement, we consider the boundary mode $\theta(\tau)$ when the singularity is at the origin of AdS$_2$.
In this case, the boundary equation simply becomes
\begin{align}
    \sum_{m>0}c_m m^2(1-m^2)\sin(m\tau) = 0\ .
\end{align}
The solution is then given by
\begin{align}\label{eq:solbdy}
    \delta \theta (\tau) = c_1 \sin(\tau)\ ,
\end{align}
where the indefinite coefficient $c_1$ is real. When the conical singularity is at the origin of AdS$_2$, the terms in the capacity of entanglement $\eqref{eq:capboundary}$ become
\begin{align}
    \begin{aligned}
    -\frac{\phi_r}{2\pi}\int^{2\pi}_0\d \tau\; R(\tau) &= \phi_r\ , \\
    \frac{\phi_r}{2\pi}\int^{2\pi}_0\d \tau\; \partial_n R(\theta)& = -\frac{\phi_r}{\pi}\int^{2\pi}_0 \d \tau \; \partial_\tau \delta\theta(\tau) = 0\ ,
    \end{aligned}
\end{align}
where we use $R(\theta) = -(\partial_\tau \theta)^2$ at $A=0$.
Finally, we get the capacity of entanglement
\begin{align}
    C = \phi_r + C_{\rm mat}\ ,
\end{align}
The boundary mode solution \eqref{eq:solbdy} comes from the ambiguity of the Schwarzian derivative.
We can fix it into $\theta(\tau) = \tau$ by using the redundancy of the conformal welding map inside the disk, which can absorb the indefiniteness of the first order perturbative equation \eqref{eq:bdyeq}.
When the gluing function is trivial ($\theta(\tau) = \tau$), the welding map is also trivial.
Then, the matter capacity is not affected by the conformal welding problem in the high temperature limit.
If we transform the coordinate $y=\frac{\beta}{2\pi}\log w$ and replace $\phi_r\to 2\pi\phi_r/\beta$ at the same time, this reproduces the island phase of the capacity of entanglement.

\section{Local analysis for dilaton}
\label{sec:local}
We check the consistency of the dilaton solution \eqref{eq:finitedilaton} on the orbifold $\CM_n$ by solving the equation of motion in two different regions, around the singularity and near the boundary of the hyperbolic disk.
We also show that the backreaction from the matter can be ignored at the leading order of $n-1$ in the high temperature limit.
In this appendix, we first set $\beta=2\pi$ for simplicity.
Later we recover $\beta$-dependence via an appropriate dimensional analysis and take the high temperature limit where the welding maps are trivialized.
In this case, the coordinate $w$ defined inside the disk $|w|\leq1$ can be extended to the outside, where $|w|\geq1$.
We concentrate on the case when the conical singularity settles down to the origin $w=0$, justified by the QES condition.

We proceed in the conformal gauge inside the disk
\begin{align}
    \d s^2 = e^{2\rho}\d w \d\bar{w}\ .
\end{align}
The equations of motion for the dilaton in the conformal gauge are
\begin{align}
    \begin{aligned}
    \label{eq:eom}
    \partial_w ^2 \phi - 2\partial_w \rho \partial_w \phi &= -2\pi T_{ww}\ ,\\
    \partial_{\bar{w}}^2\phi - 2\partial_{\bar{w}} \rho \partial_{\bar{w}} \phi &= -2\pi T_{\bar{w}\bar{w}}\ ,\\
    2\partial_w\partial_{\bar{w}}\phi-e^{2\rho}\phi &= -4\pi T_{w\bar{w}}\ ,
    \end{aligned}
\end{align}
where $T_{ab}$ in the right hand side is the matter stress tensor in the curved background.
In order to find the solution at order $n-1$, we expand the conformal factor, the dilaton and the matter stress tensor in $n-1$:
\begin{align}
    \label{eq:expand}
    \rho = \rho^{(0)} + (n-1)\,\delta \rho\ ,\qquad 
    \phi = \phi^{(0)} + (n-1)\,\delta \phi\ , \qquad
    T_{ab} = T_{ab}^{(0)} + (n-1)\,\delta T_{ab} \ .
\end{align}
Substituting \eqref{eq:expand} into \eqref{eq:eom}, we obtain the equations of order $O(n-1)$:
\begin{align}
    \begin{aligned}
    \label{eq:withsource}
    \left(\partial_w^2-2\,\partial_w \rho^{(0)} \partial_w\right)\delta \phi 
        &= -2\pi\, \delta T_{ww} +
            2\,\partial_w \delta \rho\, \partial_w \phi^{(0)} \ ,\\
    \left(\partial_{\bar{w}}^2-2\,\partial_{\bar{w}} \rho^{(0)}  \partial_{\bar{w}}\right)\delta \phi 
        &= -2\pi\, \delta T_{\bar{w}\bar{w}} + 2\,\partial_{\bar{w}}\, \delta \rho\, \partial_{\bar{w}} \phi^{(0)}\ , \\
    \left(2\,\partial_w\partial_{\bar{w}} -e^{2\rho^{(0)} }\right)\delta \phi 
        &= -4\pi\, \delta T_{w \bar{w}} +  2\,e^{2\rho^{(0)} }\delta \rho\, \phi^{(0)}\ , 
    \end{aligned}
\end{align}
where we know the solutions in the leading order \cite{Goto:2020wnk}:
\begin{align}\label{eq:solution}
    e^{2\rho^{(0)}} = \frac{4}{(1-|w|^2)^2}\ , \qquad \phi^{(0)} = \phi_r\, \frac{1+|w|^2}{1-|w|^2}\ .
\end{align}
\paragraph{Around the conical singularity}
Now we solve the equations of motion near the conical singularity $|w|\sim 0$.
First, we consider whether the matter stress tensor can be neglected in the high temperature limit.
Since the background metric is curved, the Weyl anomaly contributes to the CFT stress tensor:
\begin{align}\label{eq:weylanmly}
    -2\pi\, T_{ww} = -2\pi\, T_{ww}^{\rm flat} -\frac{c}{6}\left(\partial_w^2\rho-(\partial_w \rho)^2\right)\ ,
\end{align}
where $T_{ww}$ is the stress tensor in the curved background $\d s^2 = e^{2\rho}\,\d w \d \bar{w}$ and $T_{ww}^{\rm flat}$ is the one on flat plane $\d s^2 = \d w \d \bar{w}$. 
On the other hand, in the vicinity of the twist operator, the stress tensor in the flat space becomes
\begin{align}\label{eq:twistenegy}
    -2\pi\, T_{ww}^{\rm flat} = \frac{c}{24}\left(1-\frac{1}{n^2}\right)\frac{1}{w^2} - \left(1-\frac{1}{n}\right)\frac{\partial_A S^{\rm flat}_n|_{A=0}}{w} + O(w^0)\ ,
\end{align}
where $S_n^{\rm flat}$ represents the matter R\'enyi entropy of the interval $[A,B]$ in the flat metric and $O(w^0)$ denotes the regular term from the matter part below the order $w^0$.
Substituting \eqref{eq:twistenegy} into \eqref{eq:weylanmly}, the double pole term vanishes by the Weyl anomaly and we get the stress tensor \cite{Goto:2020wnk}:
\begin{align}
    -2\pi\, T_{ww} = -(n-1)\, \frac{\partial_A S_{\rm mat}|_{A=0}}{w} + O(w^0) + O((n-1)^2) \ .
\end{align}
The entanglement entropy $S_{\rm mat}$ contains the conformal factor $\Omega$ at $n=1$ rather than the flat metric due to the Weyl anomaly:
\begin{align}
    S_{\rm mat} = S^{\rm flat}_{\rm mat} - \frac{c}{6} \log\, \Omega(A)\ , \qquad
    \Omega(A) = \frac{1-|A|^2}{2}\ ,
\end{align}
where we define $\lim_{n\to1} S_n^{\rm flat}= S^{\rm flat}_{\rm mat}$.
The order $O(n-1)$ term of the stress tensor can be read off as
\begin{align}
    - 2\pi\, \delta T_{ww} = - \frac{\partial_A S_{\rm mat}|_{A=0}}{w} + O(w^0)\ ,
\end{align}
where we explicitly write the singular term.
Note that the stress tensor has only a simple pole term rather than double pole.
The metric near the conical singularity is given by 
\begin{align}
    \delta \rho = -\frac{1}{2}\log w\bar{w} \ .
\end{align}
Then, the right hand sides of the first two equations on \eqref{eq:withsource} become
\begin{align}\label{eq:qeslike}
    -2\pi\,\delta T_{ww}-\frac{\partial_w \phi^{(0)}}{w} \ ,
\end{align}
and similarly for $T_{\bar{w}\bar{w}}$.
Expanding the dilaton term $\partial_w \phi^{(0)}$ by $w$, the above equation has the matter part plus the dilaton term in each order of $w$. Then, we can write each term in \eqref{eq:qeslike} as
\begin{align}
    -2\pi\,\delta T_{ww} = \sum_{n=-1}^{\infty} c_n w^n\ ,\qquad
    -\frac{\partial_w \phi^{(0)}}{w} = \sum_{n=-1}^{\infty} d_n w^n\ .
\end{align}
We can consider the inverse temperature $\beta$ by introducing the coordinate $y = \frac{\beta}{2\pi}\log w$ and the replacement $\phi_r\to 2\pi\phi_r/\beta$.
After that, the dilaton terms $d_n$ are proportional to $\phi_r/\beta$. 
On the other hand, the matter terms $c_n$ are proportional to the central charge.
Therefore, at the high temperature limit $\kappa = \beta c/(24\pi \phi_r) \ll 1$, the dilaton contribution dominates the matter part in each order of $w$ in \eqref{eq:qeslike}.
This implies that the matter contributions $T_{ww}$ and $T_{\bar{w}\bar{w}}$ drop from the dilaton equations near the conical singularity at the zeroth order of high temperature expansion. 
Note that the Weyl anomaly also contributes to the diagonal part $T_{w\bar{w}}=-c\,\CR/96$ where $\CR$ is the Ricci scalar of the curved background.
Since there is a conical singularity in the orbifold, the Ricci scalar is $\CR = -2 + \frac{4\pi}{\sqrt{g}} (1-\frac{1}{n})\,\delta^2(w,\bar{w})$.
The metric is divergent at the singularity, so we eliminate the delta function and solve the dilaton around the singularity.

The right hand side is approximated at small $w$ by
\begin{align}
    \begin{aligned}
    2\,\partial_w \delta \rho\, \partial_w \phi &= - \frac{\bar{w}}{w}\frac{2\,\phi_r}{(1-|w|^2)^2}\sim -2\,\phi_r\,\frac{\bar{w}}{w}\ ,\\
    2\,\partial_{\bar{w}} \delta \rho\, \partial_{\bar{w}} \phi &= - \frac{w}{\bar{w}}\frac{2\,\phi_r}{(1-|w|^2)^2}\sim -2\,\phi_r\,\frac{w}{\bar{w}}\ ,\\
    2\,e^{2\rho}\,\delta \rho\, \phi &= -\frac{4\,\phi_r\,(1+|w|^2)}{(1-|w|^2)^3}\log |w|^2 \sim -4\,\phi_r\, \log |w|^2\ .
    \end{aligned}
\end{align}
We also have the approximated forms:
\begin{align}
    e^{2\rho} \sim 4\ ,\qquad 2\,\partial_w \rho = \frac{2\,\bar{w}}{1-|w|^2}\sim 2\,\bar{w}\ ,\qquad 2\,\partial_{\bar{w}}\rho = \frac{2\,w}{1-|w|^2}\sim 2\,w \ .
\end{align}
As a result, we get the differential equations
\begin{align}
    \begin{aligned}
    \partial_w^2 \delta \phi -2\,\bar{w}\,\partial_w\delta \phi &= -2\,\phi_r\,\frac{\bar{w}}{w}\ ,\\
    \partial_{\bar{w}}^2 \delta \phi -2\, w\,\partial_{\bar{w}}\delta \phi &= -2\,\phi_r\,\frac{w}{\bar{w}}\ ,\\
    \partial_w\partial_{\bar{w}}\delta \phi -2\, \delta \phi &= -2\,\phi_r\, \log |w|^2 \ .
    \end{aligned}
\end{align}
In order to solve the differential equation, we have to find the special solution at $|w|\sim 0$:
\begin{align}
    \delta \phi = -2\phi_r |w|^2 \log |w|^2\ .
\end{align}
This special solution vanishes at the conical singularity $w = 0$.
We can add the general solution without the source term, which is the right hand side in \eqref{eq:withsource}. Then we get the general solution near the conical singularity
\begin{align}\label{eq:nearsing}
    \delta \phi = \alpha_1\,\phi_r\,\frac{1+|w|^2}{1-|w|^2} -2\,\phi_r\,|w|^2\log |w|^2 \ ,
\end{align}
for real $\alpha_1$. 

\paragraph{Near the boundary}
We proceed to find the solution near the boundary $|w|\sim 1$.
The metric near the boundary is expanded as follows:
\begin{align}
    \d s^2 = \frac{4\,|\d w|^2}{(1-|w|^2)^2}\left[1+\frac{1}{12}\left(1-\frac{1}{n^2}\right)(1-|w|^2)^2+ \dots \right] \ .
\end{align}
The conformal factor at first order is
\begin{align}
    \delta \rho = \frac{1}{12}(1-|w|^2)^2\ .
\end{align}
Then, the equations of motion are
\begin{align}
    \begin{aligned}
    \partial_w^2\delta \phi -\frac{2\,\bar{w}}{1-|w|^2}\,\partial_w \delta \phi &= -\frac{2\,\phi_r}{3}\frac{\bar{w}^2}{1-|w|^2} \ ,\\
    \partial_{\bar{w}}^2\delta \phi -\frac{2\,w}{1-|w|^2}\,\partial_{\bar{w}} \delta \phi &= -\frac{2\,\phi_r}{3}\frac{w^2}{1-|w|^2} \ ,\\
    \partial_w\partial_{\bar{w}}\delta \phi - \frac{2}{(1-|w|^2)^2}\,\delta \phi &= \frac{2\,\phi_r}{3}\frac{1}{1-|w|^2} \ ,
    \end{aligned}
\end{align}
where we use $|w|\sim 1$.
Note that we neglect the matter stress tensors because they are regular at the boundary but the geometric contribution is divergent as shown in the right hand side of the above equations.
The special solution of these differential equations are 
\begin{align}
    \delta \phi = -\frac{\phi_r}{3}\,(1-|w|^2)\ .
\end{align}
This solution vanishes at the boundary $|w|=1$.
We can add the general solution without the source term  as in the case near the conical singularity.
Then, we have the general solution near the boundary:
\begin{align}\label{eq:nearbdy}
    \delta \phi = \alpha_2\, \phi_r\, \frac{1+|w|^2}{1-|w|^2}-\frac{\phi_r}{3}\,(1-|w|^2) \ ,
\end{align}
for real $\alpha_2$.
Now we impose the boundary condition
\begin{align}
    \delta \phi|_{\rm bdy} = 0 \ ,
\end{align}
which means $\alpha_2=0$.
However, generally, $\alpha_1 \neq \alpha_2$ because the special solution may contribute to the term proportional to $\alpha_1$.

\paragraph{Consistency check}
In the previous section, we use the conformal map $\tilde{w} = w^\frac{1}{n}$ to get the dilaton on the orbifold:
\begin{align}
    \phi = \frac{\phi_r}{n}\frac{1+|w|^\frac{2}{n}}{1-|w|^\frac{2}{n}}\ .
\end{align}
From this result, we can easily get the dilaton of the order $O(n-1)$
\begin{align}
    \delta \phi = \partial_n \phi|_{n=1} = -\phi_r\, \frac{1+|w|^2}{1-|w|^2}-\frac{2\,\phi_r\,|w|^2\log|w|^2}{(1-|w|^2)^2}\ .
\end{align}
We can check that this solution satisfies with \eqref{eq:nearsing} when $\alpha_1 = -1$ around the conical singularity and \eqref{eq:nearbdy} when $\alpha_2=0$ near the boundary.

\bibliographystyle{JHEP}
\bibliography{Replica}

\end{document}